\documentclass[10pt,journal,twoside]{IEEEtran}

\usepackage{graphicx}
\usepackage{amsmath,amssymb}
\usepackage{lastpage}
\usepackage{float}
\usepackage{subfigure}
\usepackage{afterpage}
\usepackage[usenames]{color}
\usepackage{mathrsfs}
\usepackage{upgreek}
\usepackage[hidelinks]{hyperref}
\usepackage{gensymb}
\usepackage{color}
\usepackage[dvipsnames]{xcolor}
\usepackage{tablefootnote}
\usepackage{bm}
\usepackage[capitalise]{cleveref}
\usepackage{fontawesome5}
\usepackage{multirow}
\usepackage[compress]{cite}
\usepackage{orcidlink}
\usepackage[normalem]{ulem}

\newcommand{\dd}{\mathrm{d}}
\newcommand{\obs}{\mathrm{o}}
\DeclareMathOperator{\pow}{pow}
\DeclareMathOperator{\powsqr}{square}
\DeclareMathOperator{\inv}{inv}

\DeclareRobustCommand{\svdots}{
  \vbox{%
    \baselineskip=0.33333\normalbaselineskip
    \lineskiplimit=0pt
    \hbox{.}\hbox{.}\hbox{.}%
    \kern-0.2\baselineskip
  }%
}

\begin{document}

\title{Exhaustive Symbolic Regression}

\author{Deaglan J. Bartlett\orcidlink{0000-0001-9426-7723}, Harry~Desmond\orcidlink{0000-0003-0685-9791}, and Pedro G. Ferreira\orcidlink{0000-0002-3021-2851}%
\IEEEcompsocitemizethanks{\IEEEcompsocthanksitem D.J. Bartlett is with CNRS \& Sorbonne Universit\'{e}, Institut d’Astrophysique de Paris (IAP), UMR 7095, 98 bis bd Arago, F-75014 Paris, France.\protect\\
Email: deaglan.bartlett@physics.ox.ac.uk
\IEEEcompsocthanksitem H. Desmond is with Institute of Cosmology \& Gravitation, University of Portsmouth, Dennis Sciama Building, Portsmouth, PO1 3FX, UK.\protect\\
Email: harry.desmond@port.ac.uk
\IEEEcompsocthanksitem D.J. Bartlett was with and P.G. Ferreira is with Astrophysics, University of Oxford, Denys Wilkinson Building, Keble Road, Oxford OX1 3RH, UK.}
\thanks{Manuscript received XXXX, YYYY; accepted XXXX, YYYY, date of publication XXXX, YYYY}
\thanks{(Corresponding authors: Deaglan J. Bartlett, Harry Desmond)}
\thanks{Recommended for acceptance by BBBB}
\thanks{Digital Object Identifier no. AAAA}}

\markboth{IEEE Transactions on Evolutionary Computation}%
{Bartlett \MakeLowercase{\textit{et al.}}: Exhaustive Symbolic Regression}

\maketitle

\begin{abstract}
Symbolic Regression (SR) algorithms attempt to learn analytic expressions which fit data accurately and in a highly interpretable manner. Conventional SR suffers from two fundamental issues which we address here. First, these methods search the space stochastically (typically using genetic programming) and hence do not necessarily find the best function. Second, the criteria used to select the equation optimally balancing accuracy with simplicity have been variable and subjective. To address these issues we introduce Exhaustive Symbolic Regression (ESR), which systematically and efficiently considers all possible equations---made with a given basis set of operators and up to a specified maximum complexity--- and is therefore guaranteed to find the true optimum (if parameters are perfectly optimised) and a complete function ranking subject to these constraints. We implement the minimum description length principle as a rigorous method for combining these preferences into a single objective. To illustrate the power of ESR we apply it to a catalogue of cosmic chronometers and the Pantheon+ sample of supernovae to learn the Hubble rate as a function of redshift, finding $\sim$40 functions (out of 5.2 million trial functions) that fit the data more economically than the Friedmann equation. These low-redshift data therefore do not uniquely prefer the expansion history of the standard model of cosmology. 
We make our code and full equation sets publicly available.
\end{abstract}

\IEEEdisplaynontitleabstractindextext
\IEEEpeerreviewmaketitle

\section{Introduction}
\label{sec:intro}

\IEEEPARstart{S}{ymbolic} Regression (SR) is the umbrella term for machine-learning methods that incorporate the functional form into the search space of regression, enabling optimisation over operators and their ordering as well as the values of numerical parameters.
The most popular technique for SR is genetic programming (e.g. \cite{turing, David, haupt, Lemos_2022}), in which a population of equations is evolved through several generations by means of mutations and cross-breeding. Other techniques include reinforcement learning \cite{RL_book} and physics-inspired approaches which search for symmetries to simplify the data \cite{aifeyn_0, aifeyn}.

Given a set of basis functions, the number of ways of combining such operators into equations scales exponentially with the number of operators used. The aim of SR is not just to fit the data but to be interpretable and hence produce the simplest equations possible. Increasingly complex functions achieve higher accuracy on the dataset but exhibit lower generalisability due to overfitting. Minimising complexity thus becomes a second objective, and optimisation produces a Pareto front of functions which are not surpassed in either objective without the other becoming worse. Since all models on the Pareto front are optimal with respect to these objectives, it is potentially unclear which of these should be selected if one only wanted a single equation.

By searching for simple, analytic descriptions of the data, the advantage of SR algorithms over traditional machine learning methods is their interpretability and clear (although not necessarily accurate) extrapolation behaviour when employed on data outside the range of the training set.
A potentially more important, yet more optimistic, goal of SR is to uncover ``physical laws'' from data, which are necessarily analytic. For example, SR has recently been used to rediscover the laws governing planetary motions from the trajectories of the Sun and the Solar System’s planets and large moons \cite{Lemos_2022}.
SR has also been utilised in astrophysics and beyond to find effective prescriptions for the outcomes of complex, non-linear numerical simulations. 
For example, it has been used to determine novel scaling laws \cite{Wadekar_2022}
related to the Sunyaev-Zeldovich effect \cite{Sunyaev_1972}, to improve models for the standard halo occupations distribution \cite{Delgado_2022}, and  to learn (macroscopic) transport terms required for continuity equations from microscopic-scale numerical simulations \cite{Miniati_2022}.

While state-of-the-art implementations of SR achieve high performance on test-bench problems, they are not infallible \cite{SR_review}. On any given dataset there is a (generally unknown) probability that the algorithm will fail to converge to the optimal solution, or more generally that functions near the Pareto front will not have been considered. We have therefore developed a new algorithm for SR, dubbed \emph{Exhaustive Symbolic Regression} (ESR). ESR explicitly considers every possible combination of operators from a predefined set up to a given complexity of equation, which we define as the number of nodes in the function's tree representation. By considering all possible equations, our algorithm is guaranteed to find the best-fitting function at a given complexity (provided parameters are optimised perfectly), and hence the true Pareto front. ESR differs from other deterministic approaches in explicitly evaluating every point in functional parameter space, as detailed in Sec.~\ref{sec:SR_comparison}.

Since the typical goal of SR is to find a trade-off between accuracy and simplicity to obtain the ``best'' equation for a given dataset, one must define a procedure for combining these competing metrics. Currently, there exists almost as many methods for assessing this balance as there are SR implementations. For example, with \textsc{PySR} \cite{cranmer2020discovering,pysr} one can select the most accurate equation or the equation from some region of the Pareto front with the maximum score (defined to be the negative first derivative of the logarithm of the loss function with respect to complexity). The former of these clearly has no penalty for overly complicated equations, whereas the score method has little motivation from an information-theoretic perspective. The selection criterion and complexity definition utilised in \textsc{aifeynman} \cite{aifeyn_0, aifeyn} is better motivated, drawing on the minimum description length principle. However, their final selection criterion does not consider the number of parameters used, and all parameters are penalised in the complexity by an arbitrary number (set by a precision hyperparameter), rather than by considering the constraining power of the data on these parameters. These limitations preclude the combination of accuracy and simplicity into a single objective.

To combat this fundamental issue with current SR algorithms,
we develop an implementation of the minimum description length principle to assess functions, which naturally penalises those that are more structurally or parametrically complex, and those whose parameters must be specified more precisely to yield a high accuracy.
Statistics such as the one derived in this work could
be incorporated into current and future SR implementations to enable a principled, objective model comparison: without such motivation it is difficult to assess the robustness of SR outputs.

This paper presents the first iteration of the ESR algorithm in which we consider functions of only one variable (denoted $x$ throughout) up to complexity 10. We present the algorithm in detail in \cref{sec:method,sec:MDL} and apply it to simple test cases to demonstrate its behaviour in \cref{sec:case studies}. We compare to traditional SR algorithms and discuss future avenues in \cref{sec:disc} and conclude in \cref{sec:conc}. We publicly release the results and code.

A companion paper \cite{ESR_on_RAR} presents the first detailed and substantially non-trivial application of ESR (to galaxy dynamics). 

\section{Function generation}
\label{sec:method}

In this section we describe how we generate all possible functions at a given complexity and fit the parameters of these functions to data. We then present the generated functions for an example set of basis functions and discuss the typical behaviour which can be captured by such functions.
As with all SR, the basis operator set must be specified by the analyst based on domain-specific knowledge and basic properties of the data.
Of the variant forms for a function at given complexity, we select one to be the ``unique equation'' representing that subset. While all variants of the same unique equation will give the same maximum likelihood for the data, and thus we need to only optimise the parameters of the unique equations, these variants may provide simpler descriptions of the data in the information theoretic sense that we make precise in \cref{sec:MDL}. We therefore retain all variants at this stage. We do not, however, consider any other choices for the numerical parameters unless the likelihood or dataset is changed, in which case the optimisation must be repeated.

To do this, one must have a definition of complexity. By representing the basis functions as nodes, one can represent the functions of interest as trees, and thus a natural definition of complexity is the number of nodes in the tree. We therefore wish to generate all possible expressions from our basis functions that can be represented as a tree with a given number of nodes.

In this section we outline how we achieve this. In \cref{sec:tree_representation} we demonstrate how functions can be represented as trees, which can be stored as lists of operators. We then discuss how we produce all valid equations in \cref{sec:generating_valid_trees} and how duplicates are identified and handled in \cref{sec:duplicate_checking}. The method used to optimise the parameters of the generated equations is outlined in \cref{sec:param_opt}, and we detail the number of functions produced and their properties in \cref{sec:generated_equations}.

\subsection{Tree representation}
\label{sec:tree_representation}

To begin, we will assert that all operators take a fixed number of arguments. In this paper we will consider three types of operator: binary (e.g. $+$, $\times$, $\pow$), unary (e.g. $\exp$, $\log$, $\powsqr$), or 
nullary (parameters or variables).\footnote{One could also include constants in the nullary operator list, but this would require a different operator for each constant. Parameters, on the other hand, can take any value and are optimised in a separate step (Sec.~\ref{sec:param_opt}).}
When one sees a binary operator in the tree, for example, one knows that this node is either the ``root'' of the tree or has one ``parent'' node, and that it has two ``child'' nodes (the arguments of the function). Hence, one can uniquely reconstruct the tree from a list of operators if one has a rule for traversing the tree, where the length of the list equals the complexity of the function.

\begin{figure}
	\centering
 	\includegraphics[width=\columnwidth]{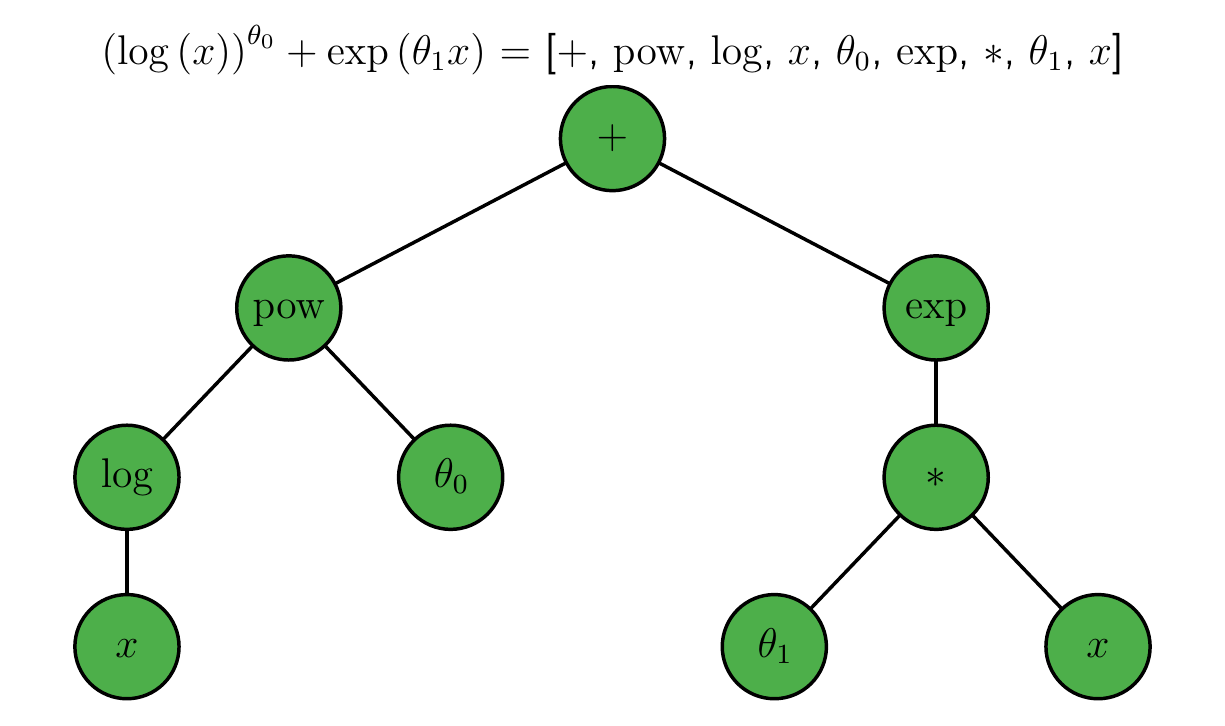}
	\caption{
	Representations of the expression
	$\left(\log\left(x\right)\right)^{\theta_0}+ \exp\left( \theta_1 x \right)$ as a tree and as a list of operators. One can generate the tree from the list using the traversal rule outlined in \cref{sec:tree_representation}. The mapping from the list to the function is unique, however a given equation does not necessarily have a unique tree representation.
  }
  	\label{fig:tree}
\end{figure}

As in Petersen et al. \cite{Petersen_2019}, the first operator in our list is the ``root'' node, which we place at the top of the tree. If this operator is binary, we draw two nodes connected to the root, and the second operator in the list is placed on the left hand node. If the operator was unary, we would only draw one node connected to the root, as the second operator labels this node. We continue traversing the tree 
in a depth-first manner
as far as possible, until we reach a nullary operator. We then go back up the tree to find the next node which has an unlabelled child on its right, and place the next operator in the list there. As an example, in \cref{fig:tree} we give one of the tree representations of the expression $\left(\log\left(x\right)\right)^{\theta_0}+ \exp\left( \theta_1 x \right)$, written both as a tree and as the corresponding list of operators.

We note two important properties of this representation which we exploit in the ESR algorithm. First, the mapping between the list and the tree is bijective, hence we choose the list representation throughout. Therefore, our task is to find all possible lists of a given length which produce valid functions. Second, a given function may have multiple tree representations and hence there is a redundancy to this notation. For efficiency in fitting parameters of the functions to data, an ESR algorithm should be able to identify these duplicates.

\subsection{Generating valid trees}
\label{sec:generating_valid_trees}

The number of equations grows exponentially as one increases complexity; for $n$ basis functions the are $n^k$ possible lists for functions comprised of $k$ nodes. However, not all of these will produce valid functions. For example, neither the list
[$\theta_0$, $+$, $x$]
nor
[$+$, $+$, $x$]
produce valid functions. Fortunately, one does not have to check all $n^k$ lists to see if they can represent functions, since e.g.
 [$*$, $x$, $\exp$] and [$*$, $x$, $\powsqr$] 
 fail for the same reason: both $\exp$ and $\powsqr$ are unary operators. 
 
 Our first task is to identify all possible tree structures (without being concerned about the node labels) and how these can be represented as lists. We produce all possible lists of length $k$ made from combinations of `0', `1' and `2' , where 0, 1 and 2 are placeholders for nullary, unary and binary operators, respectively. We then attempt to convert these lists into trees using the traversal rule outlined in \cref{sec:tree_representation}, and only keep those lists which produce valid trees. For example, for $k=4$ there are only four valid graph structures:
[1,1,1,0],
[1,2,0,0],
[2,0,1,0] and
[2,1,0,0]
out of a possible $3^4=81$ combinations of `0', `1' and `2'.
The generation of all valid functions is trivial once we have all valid tree structures: one simply considers all possible ways of decorating the nodes with the correct type of operator from the list of basis functions.
 
In fact, we do not need to consider all $3^k$ trees. Since we know that (for $k > 1$) the final node must be nullary, whereas the first node cannot be, we now only need to check the validity of $2 \times 3^{k-2}$ lists if $k >2$, whereas there is only one valid tree shape for $k=1,2$. In general this will be much smaller than $n^k$. Moreover, one does not need to consider all $2 \times 3^{k-2}$  lists, since if a list of length $\kappa < k$ nodes forms a valid function, this list cannot form the start of a list for a function of complexity $k$, so we can immediately rule out such cases.

\subsection{Duplicate checking and simplifications}
\label{sec:duplicate_checking}

Now that we have all possible equations at a given complexity for a given set of basis functions, we could stop there and fit the parameters of each equation to the data. However, this would be inefficient as there exist duplicate equations, so it is sensible to identify the unique equations and only optimise their parameters. We therefore check for the following patterns which give duplicate equations:
\begin{itemize}
	\item \textit{Tree reordering}: The trees [$+$, $\theta_0$, $x$] and [$+$, $x$, $\theta_0$] give the same equation, as we have simply swapped the left and right nodes of the `+' operator. This highlights a redundancy in the tree notation. We identify all equations which are equivalent but whose operators have a different ordering in the tree.
	\item \textit{Simplifications}: We also search for functions which are mathematically equivalent but expressed in different ways. For example, one may generate both the factorised and expanded version of a polynomial. We use the \textsc{Sympy} \cite{Sympy} package to convert our function trees into symbolic expressions to help identify these cases.
	\item \textit{Parameter permutations}: The functions $\theta_0 + \theta_1 x$ and $\theta_1 + \theta_0 x$ are distinct for given parameters $\theta_0$ and $\theta_1$, however we wish to fit these parameters to the data and thus these will return identical expressions once optimised, so should be identified as equivalent. This is a special case of reparameterisation invariance.
	\item \textit{Reparameterisation invariance}: Suppose we have two functions $\exp\left(\theta_0\right) \times x$ and $\powsqr\left(\theta_0\right) \times x$. When optimised, these will give different values of $\theta_0$, however the coefficient of $x$ in the function will be the same in both cases. Therefore, these should be considered duplicate equations, and only one of these needs to be fitted to the data.
	\item \textit{Parameter combinations}: If parameters only appear in our equations as combinations of other parameters (e.g. $\theta_0+\theta_1 + x$), then these functions can be expressed as trees of lower complexity ($\theta_0 + x$ in our example). Since the more complicated version of this function will perform worse with respect to all sensible metrics which balance simplicity with accuracy, we discard the expression with two parameters in favour of the equation with one.
\end{itemize} 

To perform this duplicate checking, in the current implementation we use the \textsc{sympy} package to transform our tree representations of functions into strings. We initially search for all unique strings, then explicitly search for patterns appearing in these equations which can be replaced with alternatives (e.g. sums, permutations, multiples, products or inverses of constants). By performing such substitutions, we are able to identify additional duplicates, and we further use \textsc{sympy}'s expand and factorisation routines to aid this search. This procedure is iterated until no new duplicates are found. We note that we do not need to find all duplicates, but only perform this search to reduce the number of equations we must fit to the data. This is the most memory intensive step of our method, as storing \textsc{sympy} representations of functions can be expensive.

Typically, one would not include integers in the set of basis function. However, some equations are more naturally expressed if one chooses to include these; for example, one would write $2x$ instead of $x+x$. We therefore additionally search for sums of equivalent expressions in the generated equations and find tree representations where these are written as a constant multiplied by the repeated sub-expression. Similarly, when evaluating logarithms and exponentials, one may prefer to write $\exp\left( 4 x\right)$ to $\powsqr \left(\powsqr \left( \exp \left( x \right) \right) \right)$. Our duplicate checker also finds such equations (and the logarithmic equivalents) and generates new trees containing multiplication by constants. By design, these equations are not unique, however these may be favourable representations when functions are compared in \cref{sec:MDL}.

\subsection{Numerical parameter optimisation}
\label{sec:param_opt}

Functions up to complexity 10 contain 0-4 free parameters $\theta$. The next step is to optimise these parameters to maximise the likelihood, $\mathcal{L}$, of the data given the function, for which we use the \texttt{BFGS} algorithm \cite{B,F}.

Any individual run of the \texttt{BFGS} optimiser may converge to a local likelihood maximum, or fail to converge at all. We therefore repeat the optimisation $N_\text{iter}$ times with different random starting positions, selecting the best result. To reduce runtime for functions that are simple to optimise, we end early if $N_\text{conv}$ iterations of the optimiser give a $\log \mathcal{L}$ within 0.5
of the best solution found so far. The count is reset to 0 if any iteration achieves a 
better $\log \mathcal{L}$
than the best solution so far found by at least 2,
which we deem to be a different local maximum. We select random starting positions from a uniform distribution within [0,3] in cases where $\theta$ itself is optimised. For the examples given in \cref{sec:case studies} we choose $N_\text{iter}=30$ and $N_\text{conv}=5$.

We have found these hyperparameter choices to result in robust optimisation in the great majority of cases: for example for the top-100 functions on the cosmic chronometer dataset (see Sec.~\ref{sec:case studies}), we find that only 1.7\% of repetitions of the optimisation procedure produce $\log(\mathcal{L}/\hat{\mathcal{L}})>1$ or $|\theta-\hat{\theta}|/\hat{\theta}>1$ (for any parameter $\theta$), where hat denotes maximum likelihood value. Excluding these, we find a standard deviation in $\log{\mathcal{L}}$ values of 0.01, which is driven by outliers because then only 2.5\% of repetitions have $\log(\mathcal{L}/\hat{\mathcal{L}})>10^{-5}$. The standard deviation of $|\theta-\hat{\theta}|/\hat{\theta}$ is 0.03 (i.e. a 3\% shift in the best-fit parameter values) while only 4.9\% of repetitions have $|\theta-\hat{\theta}|/\hat{\theta}>10^{-3}$. The optimal $N_\text{iter}$ and $N_\text{conv}$ are somewhat problem-specific, but may readily be altered by the user in our public code release.

\subsection{Generated Equations}
\label{sec:generated_equations}

\begin{figure}
    \centering
    \includegraphics[width=\columnwidth]{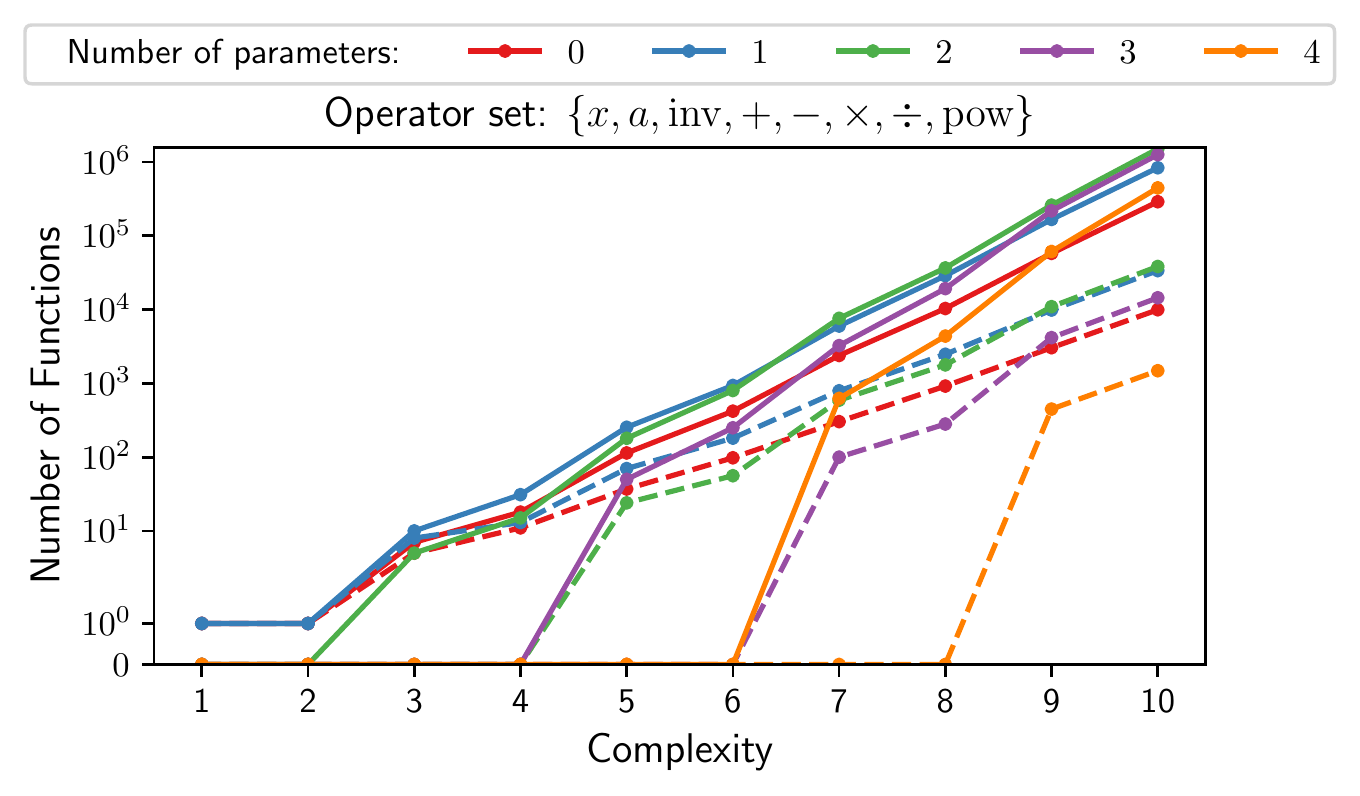}
    \caption{
    The number of trial functions containing $p$ parameters at each complexity constructed from the basis functions listed. The solid lines indicate the \emph{total} number of equations, and the dashed lines are the number of \emph{unique} equations identified in the ESR search.
    }
    \label{fig:neqs}
\end{figure}

To quantify the number of functions one should consider the impact of the simplification steps employed, we generate all possible equations for the set of basis functions  $\{x, a, \inv, +, -, \times, \div, \pow \}$ up to complexity 10.\footnote{Note that several of these operators may be constructed from others, e.g. $-$ from $+$ and $a$ and $\inv$ from $\pow$ and $a$. Including them explicitly reduces the complexity of functions containing them, increasing the diversity of equations up to a given complexity.} The number of equations with a given number of free parameters at each complexity are shown in \cref{fig:neqs}.
We show separately the number of unique equations, which are the only ones whose parameters we need to optimise, and the total number.

It is immediately clear that the simplification procedures implemented in the ESR algorithm dramatically reduces the number of functions one needs to consider. The na\"{i}ve estimate of $n^k$ functions for $n$ basis functions and $k$ nodes would suggest that one should consider approximately 168 million functions up to complexity 10. We find only 5.2 million of these are valid trees, and 134,234 of these are unique, which is a factor of 1250 smaller than the original estimate. Moreover, we only need to optimise functions which contain parameters, and thus we must run our optimisation procedure on only 119,861 functions (1400 times fewer than the original estimate).
The parameter optimisation step is readily achievable for this many functions because it is embarrassingly parallelisable: each function is optimised in isolation.

The functions generated when complexity equals 1 or 2 are extremely straightforward; at complexity 1, one can only generate a constant or the function $y=x$, and at complexity 2 one can either have a unary operator acting on $x$ or on a constant, the latter of which yields another constant. However, by complexity 3 one already has useful physical equations. For example, the function $\theta_0 / x$, which represents the Newtonian gravitational potential for a point mass orbiting a spherically symmetric body, can be expressed as [$\div$, $\theta_0$, $x$]. At complexity 5, one can fit both a straight line ([$+$, $\times$, $\theta_0$, $x$, $\theta_1$]) and a power law ([$\times$, $\theta_0$, $\pow$, $x$, $\theta_1$]).
As will be discussed further in \cref{sec:case studies}, at complexity 7 one can add a constant to a power law and thus generate a Friedmann equation for a universe consisting of a cosmological constant and a perfect fluid. A Navarro-Frenk-White (NFW) profile \cite{NFW_1997} with a free outer slope
([$\div$, $\theta_0$, $\times$, $x$, $\pow$, $+$, $x$, $\theta_1$, $\theta_2$])
is produced at complexity 9, and thus at this complexity one can generate functions which have two power law limits.
A Cauchy--Lorentz (Breit--Wigner)-like distribution is also produced at this complexity
([$\div$, $\theta_0$, $+$, $\pow$, $-$, $x$, $\theta_1$, $\theta_2$, $\theta_3$]), allowing one to fit resonances. Of course, with a different set of basis functions one would find alternative equations at these complexities; however, even with our simple basis set, it is interesting to see how many useful physical expressions can be generated by complexity 10.

Up to and including complexity 10, one does not need to optimise more than four parameters. It is straightforward to show that, if one wants an equation which depends on the input variable $x$, then at complexity $k$ any function containing nullary, unary and binary operators cannot contain more than $\lfloor (k-1)/2 \rfloor$ parameters.

The ESR algorithm is split into two distinct stages; first, one calculates all possible functions and attempts to remove duplicates, and in the second stage one fits these to the data. If one does not wish to change the basis functions, then the first step does not need to be rerun when one changes dataset or problem considered. Although we make the functions used in this work publicly available \cite{ESR_data} so one may not need to repeat the function generation, for reference, we note that generating all equations and identifying duplicates took 46 minutes at complexity 10 when run with 196 cores (dual 12-core Xeon 2.2 GHz CPUs). The computational time is significantly reduced at lower complexity; by rerunning the equation generation step for different function sets and complexities, we find that the time required at complexity $c$ scales approximately as $(1.5k)^c$ if one uses $k$ basis functions.

\section{Model selection by minimum description length}
\label{sec:MDL}

We now require a metric for quantifying the ``goodness'' of a given equation. The simplest choice is of course the likelihood, $\mathcal{L}$, itself, which describes the accuracy with which the equation reproduces the data. This is however prone to overfitting, since more complex equations will tend to match the data better while generalising worse to unseen data. 
Common statistics for tackling overfitting include the Akaike (AIC), Bayesian (BIC) and Deviance (DIC) information criteria, which combine the maximum-likelihood value with a term that depends on the number of free parameters, and the Bayesian evidence which assesses the proximity of parameters' posteriors to their priors. We do not adopt a Bayesian perspective here as we consider only the maximum-likelihood parameter values, although our procedure will naturally penalise parameters that must be fine-tuned to match the data well. Aside from the specific assumptions that go into the information criteria, their major issue is that they account for \emph{parametric} but not \emph{functional} complexity, making them suitable for numerical but not symbolic regression. Two functions with an equal number of free parameters will have the same AIC, BIC and DIC regardless of the operator set used in the functions and the number of times operators appear.

The \emph{minimum description length principle (MDL)}
provides a natural framework in which to solve this problem \cite{RISSANEN1978}. MDL posits that the best functional representation of a dataset is the one that compresses it most, so that the fewest units of information are needed to communicate the data with the help of the function. This is best understood by means of a two-part code for conveying the data: first one encodes the functional form, then one encodes the residuals of the data points around the function's expectation:
\begin{equation}
    L(D) = L(H) + L(D|H),
\end{equation}
where $L$ here denotes codelength (not to be confused with likelihood $\mathcal{L}$), $H$ is the functional hypothesis and $D$ the data.
$L(D|H)$ is an accuracy term describing how well the function fits the data, and $L(H)$ penalises more complex hypotheses. On the two extremes are \emph{i)} a function which fits the dataset perfectly so that $L(D|H)=0$ but $L(H)$ is probably very large, and \emph{ii)} the simplest hypothesis ``$0$'' which has $L(H)=0$ and $L(D|H)=L(D)$ probably very large. The best equation, minimising $L(D)$, will generally be somewhere in the middle. MDL is reviewed in \cite{MDL_review1,MDL_review2,Lanterman_2001}.

Under the Shannon--Fano coding scheme \cite{cover_thomas}, the optimal encoding of the residuals has $L(D|H)=-\log(\mathcal{L}(D|\hat{\bm \theta}))$, where ${\bm \theta}$ are the function's free parameters and a hat denotes the maximum-likelihood points (derived in \cref{sec:param_opt}).
We note that changing the units of the observable will change $L(D|H)$ by an additive constant, however, since this constant is the same for all models applied to the same data, this is unimportant in the context of model comparison.
The description length of the model, $L(H)$, is composed of a functional part and a parameter part. For the functional part, if the number of nodes in the function's tree is $k$ then we must transmit $k$ blocks of information, and each one must then specify which operator is appearing in that block. If the operator set has size $n$ (that is, the function is composed of $n$ unique operators), that requires $\log(n)$ nats per chunk, so that specifying the functional form in total requires $k\log(n)$ units of information.\footnote{Note that technically this requires the operator set for each equation to be communicated in advance, which would require some information overhead. It also uses a coding scheme that treats all operators equally, while a more efficient one would prioritise simpler or more common operators.}
We use the nat as our unit of information, so all logs, unless otherwise specified, are base-$e$.

Parameters are included in the operator set, but unlike the other operators we must also specify their values, which requires further information. Assuming that all parameters are real numbers (i.e. no simplification for integers or rationals), then the $i$\textsuperscript{th} parameter requires $\log(|\theta_i|/\Delta_i) + \log(2)$ nats, where $1/\Delta_i$ is the precision to which it is specified and the final bit ($\log(2)$ nats) specifies the sign.\footnote{In practice one would have to round up to an integer number of information units, but this is unnecessary for our purposes.} 
We use a codelength $\log(c)$ for any constant $c$ generated in a coefficient or exponent by
the function simplification process (\cref{sec:duplicate_checking}), corresponding to a natural number representation.\footnote{Negation must therefore be included in the operator list should any constant be negative.}
We note that this represents a lower bound on the number of nats required to transfer $\theta_i$ or $c$, since, to be decipherable, the codewords used must form a prefix set. Various optimal schemes are possible, which require a code length $L(c)$ to transfer the natural number $c$ such that $\log (c) < L(c) < \log (c) + R(c)$, where $R(c)/\log(c) \to 0$ as $c \to \infty$ \cite{Rissanen_1983}, but differ in their specific values for a given $c$. Since all optimal solutions have $\log (c)$ as their leading order term \cite{Rissanen_1983}, we choose to ignore higher order corrections to remain agnostic to the exact prefix code utilised. As such, this term penalises parameters slightly less than it would need to in practice.

The total description length is then
\begin{align}\label{eq:length}
    L(D) & \equiv L(D|H) + L(H) \\
    & = -\log(\mathcal{L}(\hat{{\bm \theta}})) + k\log(n) + p\log(2) + \sum_j \log(c_j) \nonumber \\
    &  \quad + \sum_i^p \log(|\hat{\theta}_i|/\Delta_i), \nonumber
\end{align}
where $p$ is the total number of free parameters.
The only degrees of freedom here are the $\Delta_i$. In keeping with the MDL principle, we choose these by minimising the right hand side of \cref{eq:length} \cite{RISSANEN1978}. Increasing $\Delta_i$ (i.e. specifying the parameter to lower precision) reduces the parameter codelength $\log(|\theta_i|/\Delta_i)$ but increases the  first, log-likelihood term because the communicated, rounded value is likely to be further from the maximum-likelihood value. Rounding to the nearest $\Delta_i$ gives $\theta_i$ a uniform probability in [$\hat{\theta}_i-\Delta_i/2, \hat{\theta}_i+\Delta_i/2$]. We find what loss this incurs by Taylor-expanding $-\log(\mathcal{L})$ around its minimum:
\begin{equation}
    -\log(\mathcal{L}(\mathbf{\hat{\bm \theta}} + \mathbf{d})) \approx -\log(\mathcal{L}(\mathbf{\hat{\bm \theta}})) + \frac{1}{2} \mathbf{d}^{\rm T} \: \mathbf{\mathsf{I}} \: \mathbf{d},
\end{equation}
where $d_i$ is a uniform random variable in [$-\Delta_i/2, \Delta_i/2$] and $\mathbf{\mathsf{I}}$ is the Fisher information matrix.
This provides an expected contribution to the description length
\begin{equation}
    L(\mathbf{\Delta}) = \frac{1}{2} \sum_{ij} \langle \mathsf{I}_{ij} d_i d_j \rangle - \sum_i \log(\Delta_i),
\end{equation}
where the 2\textsuperscript{nd} term on the right hand side comes from the denominator of the final term in \cref{eq:length} (the only part that depends on $\Delta_i$). Since the $d_i$ are independent of one another, $\langle \mathsf{I}_{ij} d_i d_j \rangle = \mathsf{I}_{ii} \langle d_i^2 \rangle \delta_{ij}$ and we see that only the diagonal elements of the Fisher matrix contribute and the parameters' contributions decouple. Using that $\langle d_i^2 \rangle = \Delta_i^2/12$, we have
\begin{equation}
    L(\Delta_i) = \frac{1}{24} \mathsf{I}_{ii} \Delta_i^2 - \log(\Delta_i),
\end{equation}
which is minimised for $ \Delta_i = \left({12}/{\mathsf{I}_{ii}}\right)^{1/2}$.
We choose this $\Delta_i$ such that the expectation value of $L(\mathbf{\Delta})$ is minimised \cite{Wallace_1968,RISSANEN1978}, but note that an alternative common choice is to minimise the maximum of this quantity \cite{Rissanen_1983}.
This difference---corresponding to minimisation of an average or worse-case description length over possible datasets---is not important for our purposes.
This implies a total description length
\begin{align}\label{eq:length_final}
    L(D) = 
    &-\log(\mathcal{L}(\hat{\mathbf{\theta}})) + k\log(n) - \frac{p}{2} \log(3)\\
    &+ \sum_j \log(c_j) + \sum_i^p \left( \frac{1}{2}\log(\mathsf{I}_{ii}) + \log(|\hat{\theta}_i|) \right), \nonumber
\end{align}
which is the quantity we seek to minimise over the function set.\footnote{Note that we optimise the parameter vector purely for the likelihood. This is in general not the same as optimising it for the codelength due to the $\frac{1}{2}\log(\mathsf{I}_{ii}) + \log(|\hat{\theta}_i|)$ term. That alternative procedure is more computationally expensive as the full parameter codelength would have to be calculated for each step of the optimiser.}

There may arise cases in which $|\hat{\theta}_i|/\Delta_i < 1$ so that the maximum-likelihood value of a parameter cannot be distinguished from 0 within the tolerance given by $\Delta_i$. Following Rissanen \cite{RISSANEN1978}, in this case we set the parameter to 0, recalculate the likelihood and reduce $p$ by 1. Since this occurs precisely in cases where the parameter is poorly distinguished from 0 anyway, this generally has little effect on $L(D)$.

When we assess the tradeoff between simplicity and accuracy, the true tree representation will be important. Instead of simply discarding all duplicates found in \cref{sec:duplicate_checking}, we determine the function which maps the parameters in the unique function, $\bm{\varphi}$, to those in the original equation, $\bm{\theta}$. For example, equations $\varphi_0 x$ and $x / \theta_0$ are viewed as duplicate equations, so we only need to optimise the first of these. One can then use $\theta_0 (\varphi_0) = 1 / \varphi_0$ to obtain the optimal parameter value for the second equation.

During the optimisation step, we compute the observed Fisher matrix via finite differences with the \textsc{numdifftools} package \cite{numdifftools}. This is done for the unique functions at their maximum likelihood points $\bm{\varphi}=\bm{\hat{\varphi}}$, which, by the chain rule, can be related to a different set of parameters, $\bm{\theta}$, as
\begin{equation}
    \begin{split}
        \bm{\mathsf{I}}_{ij}^{\left(\bm{\varphi}\right)} 
	    &\equiv 
	    \left. - \frac{\partial^2 \log \mathcal{L}}{\partial \varphi_i \partial \varphi_j} \right|_{\bm{\varphi}=\hat{\bm{\varphi}}} \\
	    &=
	    \left( - \frac{\partial \theta_\ell}{\partial \varphi_i}
	    \frac{\partial \theta_k}{\partial \varphi_j}
	    \frac{\partial^2 \log \mathcal{L}}{\partial \theta_k \partial \theta_\ell}
	    - \frac{\partial^2 \theta_k}{\partial \varphi_i \partial \varphi_j} \frac{\partial \log \mathcal{L}}{\partial \theta_k}
	     \right)_{\bm{\theta}=\hat{\bm{\theta}}},
    \end{split}
\end{equation}
where we use the Einstein summation convention. The second term is zero by definition at the maximum likelihood point, hence, defining 
the Jacobian for our transformation to be $\bm{\mathsf{J}}_{ij} \equiv \partial \theta_i / \partial \varphi_j $
we can transform the Fisher matrix as
\begin{equation}
	\bm{\mathsf{I}}^{\left(\bm{\theta}\right)}_{ij} 
	\equiv 
	\left. - \frac{\partial^2 \log \mathcal{L}}{\partial \theta_i \partial \theta_j} \right|_{\bm{\theta}=\hat{\bm{\theta}}}
	=
	\left[ \left( \bm{\mathsf{J}}^{-1}\right)^{\rm T}
	\bm{\mathsf{I}}^{\left(\bm{\varphi}\right)}
	\bm{\mathsf{J}}^{-1}
	\right]_{ij}.
\end{equation}
For the previous example, this means that $\bm{\mathsf{I}}$ for the expression $x / \theta_0$ is equal to $\bm{\mathsf{I}}$ for $\varphi_0 x$ multiplied by $\hat{\varphi}_0^4$.

MDL is not wedded to the notion that there exists a ``true'' hypothesis which it is the job of the model selection algorithm to find. Rather,
the principle
formalises Occam's razor by seeking to find the most economical description of the data, which is typically the one that generalises the best. Thus, it is naturally expected that MDL will prefer less simple but more accurate functions the more data is available: the simplicity terms in \cref{eq:length_final} become less important compared to the likelihood.
This makes model selection based on MDL conservative, as one should only hope to recover the generating equation when data of sufficient quality and quantity are available; otherwise one prefers simpler representations more likely to have reasonable extrapolation behaviour. The accuracy and simplicity objectives of SR are made commensurable by expressing them in terms of bits of information, allowing single-objective optimisation and the collapse of the Pareto front into the globally optimal solution. If, despite the advantages of MDL, an alternative quantification of Occam's razor is desired, this may be readily calculated from the parameter-optimised full function set if it depends only on quantities appearing in Eq.~\ref{eq:length_final}. As an example, the BIC may be derived as a special case of the description length \cite{MDL_review2}.

\section{Case Study -- The cosmic expansion rate}
\label{sec:case studies}

As an example of an application of the ESR algorithm, in this section we attempt to determine the expansion rate of the Universe near the present-day. In the concordance model of cosmology (called $\Lambda$CDM for $\Lambda$ [dark energy] and Cold Dark Matter), the universe is homogeneous, isotropic and spatially flat on large scales and at late times is composed primarily of dark matter and dark energy.
The expansion of the Universe is completely determined by the Friedmann equation,
\begin{equation}
    \label{eq:Friedmann standard form}
    H(a)^2 = H_0^2 (\Omega_\Lambda + \Omega_m a^{-3}),
\end{equation}
where $a$ is the scale factor of the Universe, $H \equiv \dot{a}/a$ is the ``Hubble parameter'' describing the expansion rate, $H_0$ is $H$ evaluated at the present time ($a=1$) and $\Omega_X = \frac{8 \pi G \rho_X}{3 H_0^2}$ is a normalised density of component $X$ at the present day. $a$ can be equivalently expressed in terms of redshift $z$ describing the fractional increase in the wavelength of light emitted at that scale factor due to the expansion of the universe as it travels to us: $a \equiv 1/(1+z)$. Defining
$y \left(x \equiv 1 + z \right) \equiv \left( H\left( z \right) \right)^2$, the $\Lambda$CDM expansion rate can be written
\begin{equation}
    \label{eq:Friedmann_LCDM}
    y_{\rm \Lambda CDM} \left(x \right) = \theta_0 +  \theta_1 x^3,
\end{equation}
with $\theta_0 = \frac{8 \pi G \rho_\Lambda}{3}$ and $\theta_1 = \frac{8 \pi G \rho_\text{dm}}{3}$. We can also consider a generalisation (``$\Lambda$fluid'') in which the matter component has an arbitrary dependence on the scale factor, implying 
\begin{equation}
    \label{eq:Friedmann free w}
    y_{\rm \Lambda fluid}\left(x\right) = \theta_0 + \theta_1 x^{\theta_2}.
\end{equation}
For the basis functions described in \cref{sec:generated_equations} these are complexity-7 functions ($+$, $\theta_0$, $\times$, $\theta_1$, $\pow$, $x$, $3$ or $\theta_2$) and hence well within the scope of ESR.

We apply the ESR algorithm using this basis separately to a sample of $H(z)$ measurements from cosmic chronometers and to measurements of the distances to supernovae of known redshift to determine whether \cref{eq:Friedmann_LCDM} is preferred by the data, or whether there exists alternative best-fitting functions. We 
choose a maximum complexity of 10 and a basis operator set $\{x, a, \inv, +, -, \times, \div, \pow \}$. 
We outline the observational data used and the likelihoods employed in \cref{sec:observations}, and the results are presented and discussed in \cref{sec:results}.
In \cref{sec:mock} we analyse mock cosmic chronometer data generated according to $\Lambda$CDM, to determine the requirements for the MDL principle to select this as the best model.
We note that SR has previously been applied to cosmological data using genetic programming
\cite{Bogdanos_2009,Nesseris_2010,Nesseris_2012,Sapone_2014,Arjona_2020a,Arjona_2020b,Arjona_2020c,Alestas_2022}.

\subsection{Observational data}
\label{sec:observations}

\subsubsection{Cosmic chronometers}
\label{sec:cosmic_chronometers}

Cosmic chronometers \cite{Jimenez_2002} use stellar populations that evolve passively as standard clocks to constraint $H(z)$.
In this section we use a sample of 32 cosmic chronometer measurements of $H\left(z\right)$ from the literature \cite{Moresco_2015,Moresco_2016,Ratsimbazafy_2017,Stern_2010,Simon_2005,Borghi_2022,Zhang_2014,Moresco_2012}, compiled by  Moresco et al. \cite{Moresco_2022}. We treat each of these measurements as independent and compare to our prediction, 
$H\left(z \right) = \sqrt{y}$,
by assuming a Gaussian likelihood,
\begin{equation}
    \label{eq:cc_likelihood}
    \log \mathcal{L} = 
    - \frac{1}{2} \sum_i
    \left(
    \frac{\left( H \left(z_i^\obs \right) - H_i^\obs \right)^2}{\sigma_i^2}
    + \log \left( 2 \pi \sigma_i^2 \right)
    \right),
\end{equation}
where $\{H_i^\obs\}$ are the set of observed Hubble rates at redshifts $\{z_i^\obs\}$, each of which has an associated uncertainty $\sigma_i$. For simplicity, we will ignore off diagonal terms in the covariance (although see \cite{Moresco_2020}).

\subsubsection{Type Ia Supernovae}
\label{sec:supernovae}

Type Ia
supernovae were the first objects to provide evidence of an accelerating Universe \cite{Riess_1998,Perlmutter_1999}, and have thus been used extensively over the last few decades to constrain parameters of $\Lambda$CDM and its extensions. 
In this work we utilise the Pantheon+ sample of 1590 supernovae \cite{Scolnic_2021} alongside the Cepheid distances provided by the SH0ES collaboration \cite{SH0ES_2022}. We make use of the publicly available (non-diagonal) covariance matrix, $\mathbf{\mathsf{\Sigma}}$, produced as part of the Pantheon+ cosmological analysis \cite{Pantheon_2022}.

To convert the Hubble parameter we are learning, 
$H\left(z \right) = \sqrt{y}$,
to a distance modulus, $\mu_i^\obs$, we will assume a flat universe. The luminosity distance, $d_{\rm L}$, to a supernova at redshift $z$ is then
\begin{equation}
    d_{\rm L} \left( z \right) =
    \left( 1 + z \right)
    \int_0^z \frac{c \, \dd z^\prime}{H \left( z^\prime \right)},
\end{equation}
where $c$ is the speed of light, and the corresponding distance modulus is
$\mu \left( z \right) = 5 \log_{10} \left( {d_{\rm L} \left( z \right)}/{10 {\rm \, parsec}} \right)$ and parsec is an astronomical unit of distance.
This is then compared to the observed supernova data using a Gaussian likelihood
\begin{equation}
    \begin{split}
        \log \mathcal{L} = 
        &-\frac{1}{2}
        \left(\bm{\mu} \left( \bm{z}^\obs \right) - \bm{\mu}^\obs  \right)^{\rm T}
        \mathbf{\mathsf{\Sigma}}^{-1}
        \left(\bm{\mu} \left( \bm{z}^\obs \right) - \bm{\mu}^\obs \right) \\
        & - \frac{1}{2}\log \left( \det \left( 2 \pi \mathbf{\mathsf{\Sigma}} \right) \right),
    \end{split}
\end{equation}
where $\bm{\mu}^\obs$ and $\bm{\mu} \left( \bm{z}^\obs \right)$ are vectors describing the set of $\mu^\obs, \mu\left(z^\obs\right)$ for a given supernova. Since $\mathbf{\mathsf{\Sigma}}$ is independent of $y\left(x\right)$, we do not include the final term in our optimisation as this adds an arbitrary constant to $\log\mathcal{L}$, and thus to $L\left(D\right)$.

\subsection{Results}
\label{sec:results}

\cref{tab:cc_results,tab:panth_results} show the best-fit ESR equations for the cosmic chronometer and Pantheon+ samples respectively. In both cases, we observe that the top five functions---ranked according to their description lengths---contain only one or two parameters, since functions with more parameters require more information to specify their values and they require trees of higher complexity. For reference, in these tables we also show the first appearance of the Friedmann equations for cold dark matter (\cref{eq:Friedmann_LCDM}) and the more general $\Lambda$fluid (\cref{eq:Friedmann free w}).
We find that the solution with a free exponent is approximately $\Lambda$CDM for both datasets, since $\theta_2 \approx 3$. For both datasets we find that \cref{eq:Friedmann free w} is not preferred by the data, but ranks in the top 100 functions, whereas $\Lambda$CDM performs better, and is in the top 40 functions.
The $\Lambda$CDM equation ranks slightly higher for the Pantheon+ data owing to the increased constraining power of the data: the change in $\log\mathcal{L}$ for the Pantheon data between \cref{eq:Friedmann free w} and the MDL function is smaller than for cosmic chronometers.
Since constants are not included in the operator set this function was generated here from the multiplication of three $x$s and hence has complexity 9, rather than 7 which would be the case if it were formed from $\pow\left(x,3\right)$. If we replace the complexity 9 variant by the complexity 7 one, then the description length of the $\Lambda$CDM function marginally decreases, improving
the function's ranking to 38\textsuperscript{th} and 34\textsuperscript{th} for the cosmic chronometer and Pantheon+ data respectively. Although \cref{eq:Friedmann free w} has a larger $\log\mathcal{L}$ than any of the top-5 functions for the Pantheon+ data, there are 201 functions which perform better than this equation when ranked by likelihood.

Given that we find that a two-component universe is not the most economical description of the data, one may wonder whether a single-component universe provides a better fit. Indeed, the highest ranked function for the cosmic chronometer data is proportional to the square of $1+z$. This would correspond to a universe consisting exclusively of curvature, however a more appropriate interpretation is that this exponent falls between the values of dark matter (a cubic in $1+z$) and dark energy (a constant). Similarly, we find that a single component universe, $y = \theta_0 x^{\theta_1}$, is the 11\textsuperscript{th} highest-ranked function for the Pantheon+ data, with $\theta_1=1.42$. Again, this exponent is presumably capturing an ``average'' contribution of the two terms in \cref{eq:Friedmann free w}, but it provides a worse fit by a change in log-likelihood of
4.4.
Interestingly, although containing just a single parameter, we find that a de Sitter universe ($y = \theta_0$) is severely disfavoured by the data; this expansion history ranks 10,058\textsuperscript{th} for the cosmic chronometer data and our optimiser fails 
to converge on a value of $\theta_0$ for the Pantheon+ sample.

We see that \cref{eq:Friedmann free w} has a better accuracy (smaller $-\log\mathcal{L}$) than any of the top-5 functions, when ranked by description length. If instead we order \textit{all} functions by log-likelihood, one finds that \cref{eq:Friedmann free w} actually performs worse;
it is the 829\textsuperscript{th} and 202\textsuperscript{nd} most accurate function for the cosmic chronometer and Pantheon+ data, respectively.
The pure $\Lambda$CDM expression (\cref{eq:Friedmann_LCDM}) is ranked even lower; it could not have performed better as it is a special case of \cref{eq:Friedmann free w}.
This highlights the importance of looking at functions off the Pareto front in SR: the ``truth'' (or at least the function for which we have the strongest theoretical prior) may not be a Pareto-optimal solution.

\begin{figure}
    \centering
    \includegraphics[width=\columnwidth]{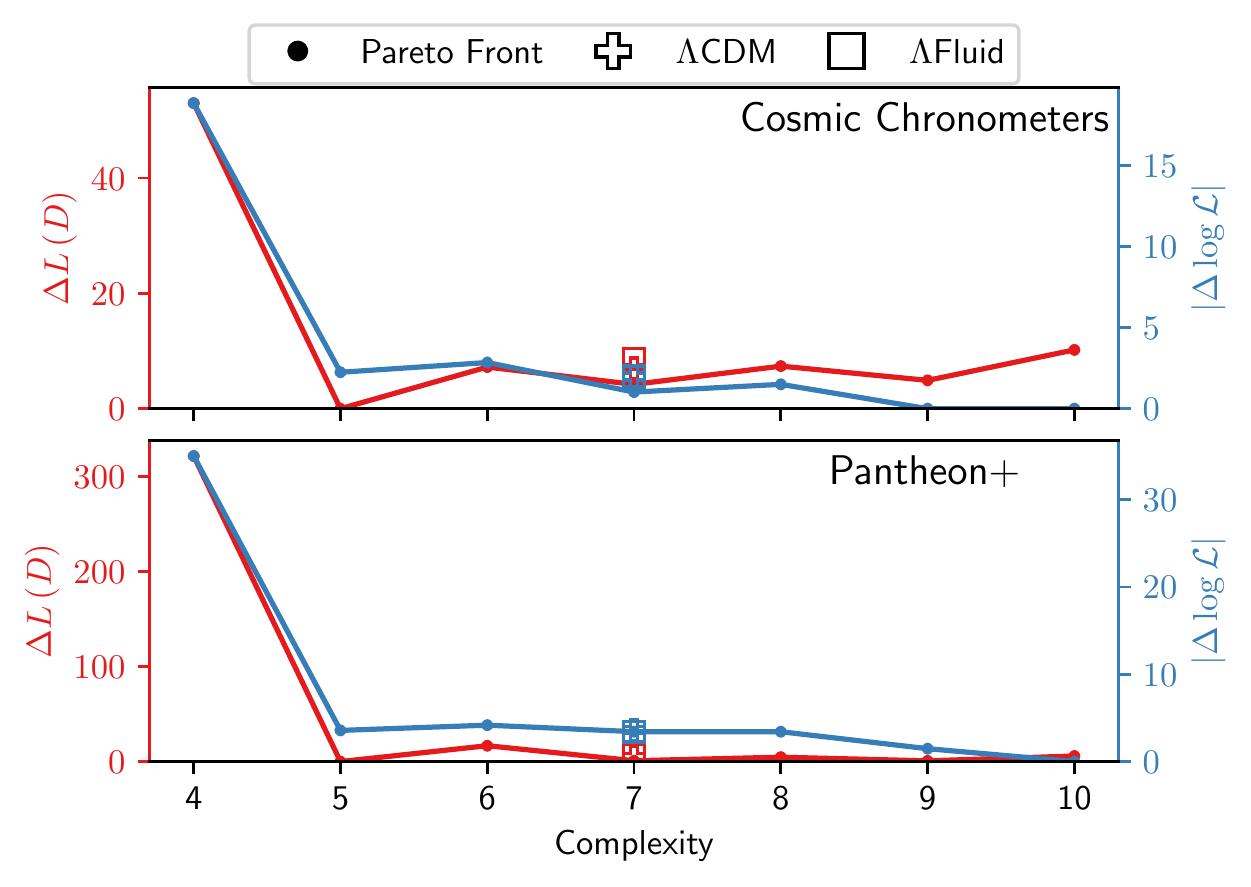}
    \caption{
    Pareto front of functions generated with the ESR algorithm, compared to the cosmic chronometer (upper) and Pantheon+ (lower) data. We show the best-fitting functions according the change in the description length, $L$, and the likelihood, $\mathcal{L}$, relative to the corresponding minima, alongside the best-fit $\Lambda$CDM solutions and a more general Friedmann equation (\cref{eq:Friedmann free w}).
    }
    \label{fig:pareto}
\end{figure}

When one assesses equations according to the likelihood, the five best functions have complexities of either 9 or 10. For example, the highest likelihood function for the Pantheon+ data is
\begin{equation}
    y (x) = \left( \theta_1 + \frac{1}{\theta_0 - x} \right)^{\theta_2^x},
\end{equation}
with $\bm{\theta} = (1.87, 1788.46, 1.15)$. This is significantly more complex than the functions selected by MDL and arguably harder to interpret, demonstrating the advantage of using MDL to choose simple yet accurate functions. Note that this problem is exacerbated at higher complexity: one can always produce either a better-fitting, or at least equally well-fitting, function by adding numbers near 0 or multiplying by numbers near 1, increasing the complexity in steps of 2.
As one could anticipate, by making functions arbitrarily complex one can improve the fit, however selecting functions according the MDL gives a principled and unambiguous way of determining what complexity of expression is warranted by the data. This can be observed in \cref{fig:pareto}, where we plot the Pareto fronts of the cosmic chronometer and Pantheon+ data, comparing both the likelihood and description lengths. As one increases complexity, the maximum-likelihood plateaus to become approximately constant, whereas the description length shows a well-defined minimum at a complexity of 5 for both datasets. By collapsing the Pareto front into a single number (the description length), the task of choosing the best function becomes well-defined.

\begin{figure}
    \centering
    \includegraphics[width=\columnwidth]{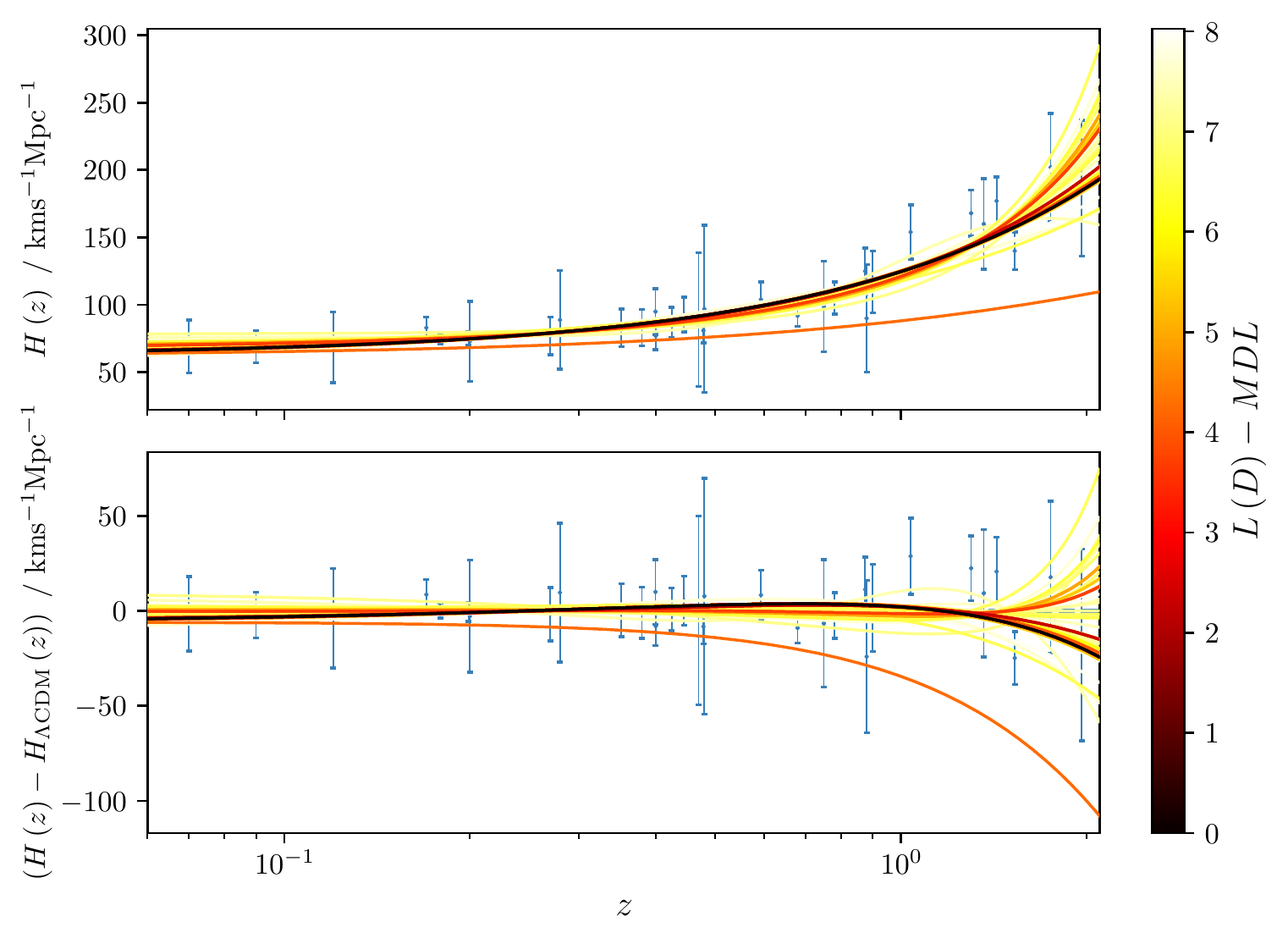}
    \caption{Hubble parameter, $H$, as a function of redshift, $z$, learned using the ESR algorithm from cosmic chronometer data (blue points).
    In the bottom panel we subtract the best-fit $\Lambda$CDM prediction. We plot the top 150 functions up to complexity 10, ranked and coloured by their description length, $L\left(D\right)$, relative to the minimum, MDL.
    }
    \label{fig:cc_prediction}
\end{figure}

\begin{figure}
    \centering
    \includegraphics[width=\columnwidth]{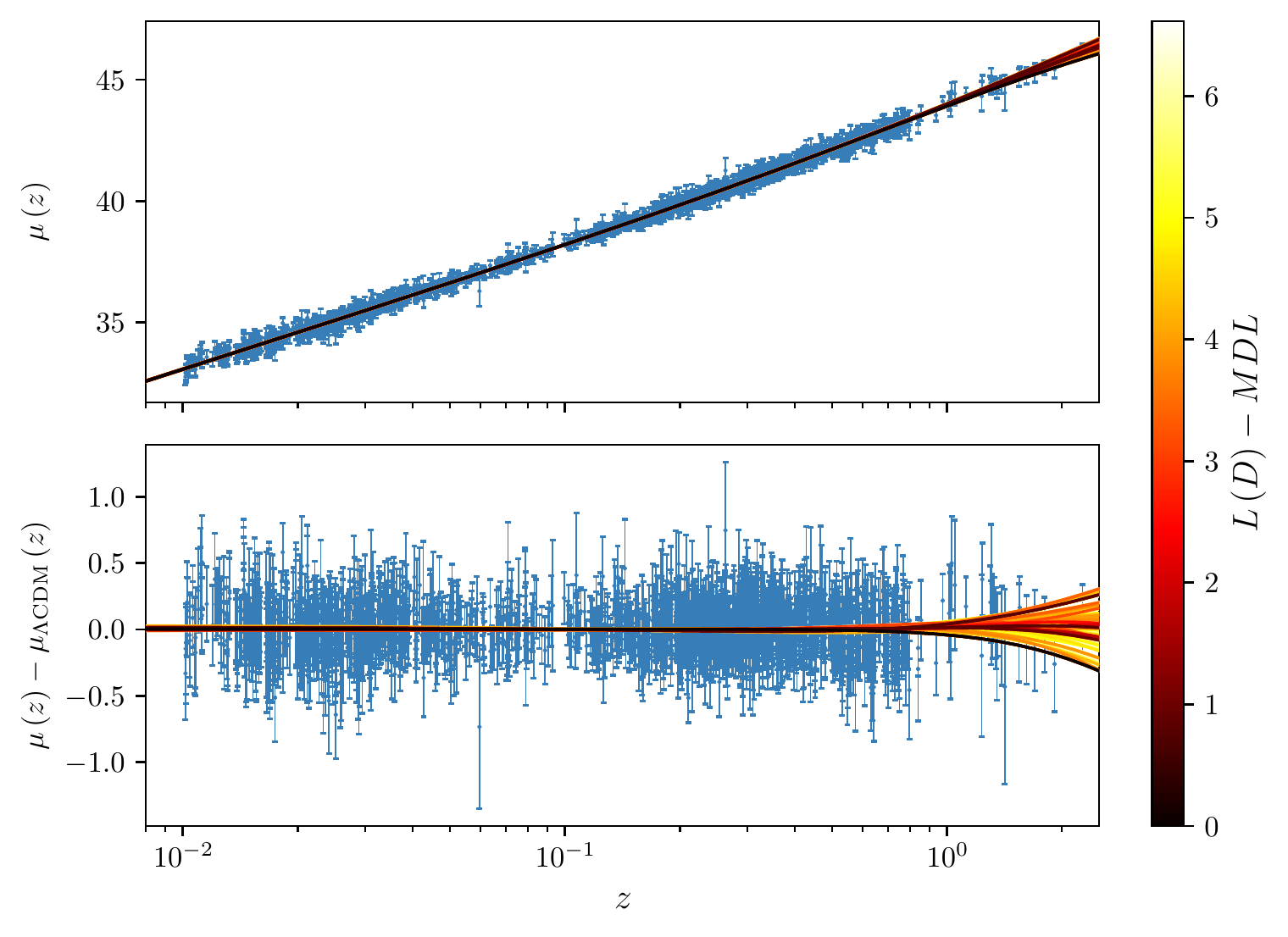}
    \caption{Distance moduli, $\mu$, to Type Ia supernovae in the Pantheon+ catalogue (blue points) as a function of redshift, $z$. We plot the 150 highest ranked functions up to complexity 10 inferred using the ESR algorithm and in the lower panel we plot the residuals relative to the best-fit $\Lambda$CDM prediction.
    }
    \label{fig:pantheon_prediction}
\end{figure}

To gain intuition as to why equations which are not Friedmann-like are preferred, in \cref{fig:cc_prediction,fig:pantheon_prediction} we plot the 150 functions with the smallest description lengths alongside the data they are fitted to, and show the residuals with respect to $\Lambda$CDM. Almost all functions are effectively indistinguishable for $z\lesssim1$, where the majority of the data is. To understand this, let us first 
expand \cref{eq:Friedmann standard form}, using $\Omega_{\rm m0} = 1 - \Omega_{\rm \Lambda 0}$
\begin{equation}
    y_{\rm \Lambda CDM} \left(x\right)
    = H_0^2 \left( 1 + 3 \Omega_{\rm m 0}z + 3 \Omega_{\rm m 0}z^2 + \Omega_{\rm m 0}z^3 \right).
\end{equation}
If we now expand the highest ranked function from the Pantheon+ data about $z=0$,
\begin{equation}
    y \left( x \right) = \theta_0 x^x = \theta_0 \left( 1 + z + z^2  + \frac{1}{2}z^3 + \mathcal{O}\left(z^4\right) \right),
\end{equation}
one sees that, since $\Omega_{\rm m 0} = 0.315 \pm 0.007\approx 1/3$ when inferred using the cosmic microwave background \cite{Planck18_Parameters}, the Taylor series for the $\Lambda$CDM and best-fitting function are approximately equal up to and including second order, if one identifies $\theta_0 = H_0^2$. It is perhaps, therefore, merely a coincidence that a one-parameter function can fit the local expansion rate, owing to dark matter comprising approximately one third of the present-day energy budget of the Universe and having an equation of state parameter $w=0$.

At redshifts $z\gtrsim 1$, we see that the predicted Hubble rate and distance moduli diverge rapidly for the different functions, due to the lack of data in this regime. By having an analytic expression, one can easily identify how these equations will diverge. One could therefore combine this analysis with higher-redshift probes to more tightly constrain the true Hubble rate, since we will no longer be in a regime where the Taylor expansion above is valid, and thus the functions identified in this work will have very different predictions.

\begin{table*}
    \begin{center}
		\begin{tabular}{|c|c|c|c|c|c|c|c|c|c|c}
		\hline
		\multirow{2}{*}{Rank} & \multirow{2}{*}{$y\left( x \right) \ / \ {\rm km^2 s^{-2} Mpc^{-2}}$} &
		\multirow{2}{*}{Complexity} & \multicolumn{3}{c|}{Parameters} & \multicolumn{4}{c|}{Description length} \\
		\cline{4-10}
		& &  & $\theta_{0}$ & $\theta_{1}$ & $\theta_{2}$& Residuals$^1$ & Function$^2$ & Parameter$^3$ & Total \\
		\hline
			1 & $\theta_{0} x^{2}$ & 5 & 3883.44 & - & -  & 8.36 & 5.49 & 2.53 & 16.39 \\
			2 & $\left|{\theta_{0}}\right|^{x^{\theta_{1}}}$ & 5 & 3982.43 & 0.22 & -  & 7.97 & 5.49 & 5.24 & 18.70 \\
			3 & $\theta_{0} \left|{\theta_{1}}\right|^{- x}$ & 5 & 1414.43 & 0.31 & -  & 7.57 & 6.93 & 5.58 & 20.08 \\
			4 & $\theta_{0} x^{\theta_{1}}$ & 5 & 3834.51 & 2.03 & -  & 8.35 & 6.93 & 5.08 & 20.36 \\
			5 & $x^{2} \left(\theta_{0} + x\right)$ & 7 & 3881.85 & - & -  & 8.36 & 9.70 & 2.53 & 20.60 \\
\svdots & \svdots & \svdots & \svdots & \svdots & \svdots & \svdots & \svdots & \svdots & \svdots \\
			39 & $\theta_{0} + \theta_{1} x^{3}$ & 9 & 3164.02 & 1481.71 & -  & 7.28 & 12.48 & 3.76 & 23.51 \\
\svdots & \svdots & \svdots & \svdots & \svdots & \svdots & \svdots & \svdots & \svdots & \svdots \\
			84 & $\theta_{0} + \theta_{1} x^{\theta_{2}}$ & 7 & 3322.96 & 1374.97 & 3.08  & 7.27 & 11.27 & 6.52 & 25.06 \\
		\hline
		\end{tabular}
		\begin{tabular}{c}
			 $^1 - \log\mathcal{L} ( \hat{\bm{\theta}} ) $ \qquad\qquad $^2 k\log(n) + \sum_j \log(c_j)$ \qquad\qquad $^3 - \frac{p}{2} \log(3) + \sum_i^p (\frac{1}{2}\log(I_{ii}) + \log(|\hat{\theta}_i|))$ \
		\end{tabular}
    \caption{Highest ranked functions for
    $y\left(x\equiv1+z\right)=\left(H\left(z\right)\right)^2$
    inferred from cosmic chronometers using the ESR algorithm.
    The functions are ordered by their description length, $L\left(D\right)$.
    The quoted parameter values are those that maximise the likelihood.
    }
    \label{tab:cc_results}
    \end{center}
\end{table*}

\begin{table*}
    \begin{center}
       \begin{tabular}{|c|c|c|c|c|c|c|c|c|c|c}
		\hline
		\multirow{2}{*}{Rank} & \multirow{2}{*}{$y\left( x \right) \ / \ {\rm km^2 s^{-2} Mpc^{-2}}$} & \multirow{2}{*}{Complexity} & \multicolumn{3}{c|}{Parameters} & \multicolumn{4}{c|}{Description length} \\
        \cline{4-10}
		& &  & $\theta_{0}$ & $\theta_{1}$ & $\theta_{2}$& Residuals$^1$ & Function$^2$ & Parameter$^3$ & Total \\
		\hline
			1 & $\theta_{0} x^{x}$ & 5 & 5345.02 & - & -  & 706.18 & 6.93 & 5.11 & 718.22 \\
			2 & $\left|{\theta_{0}}\right|^{x^{\theta_{1}}}$ & 9 & 5280.11 & 0.16 & -  & 705.11 & 5.49 & 8.41 & 719.01 \\
			3 & $\theta_{0} \left|{\theta_{1}}\right|^{- x}$ & 5 & 1694.95 & 0.32 & -  & 701.79 & 6.93 & 10.33 & 719.05 \\
			4 & $\theta_{0} x^{x^{\theta_{1}}}$ & 7 & 5378.69 & 0.78 & -  & 702.45 & 9.70 & 6.98 & 719.13 \\
			5 & $\left|{\theta_{0}}\right|^{\left|{\theta_{1}}\right|^{x}}$ & 5 & 1898.47 & 1.14 & -  & 701.88 & 5.49 & 12.64 & 720.02 \\
\svdots & \svdots & \svdots & \svdots & \svdots & \svdots & \svdots & \svdots & \svdots & \svdots \\
			37 & $\theta_{0} + \theta_{1} x^{3}$ & 9 & 3591.09 & 1773.63 & -  & 701.85 & 12.48 & 8.81 & 723.13 \\
\svdots & \svdots & \svdots & \svdots & \svdots & \svdots & \svdots & \svdots & \svdots & \svdots \\
			96 & $\theta_{0} + \theta_{1} x^{\theta_{2}}$ & 7 & 3280.83 & 2069.32 & 2.73  & 701.64 & 11.27 & 12.19 & 725.10 \\
		\hline
		\end{tabular}
		\begin{tabular}{c}
			 $^1 - \log\mathcal{L} ( \hat{\bm{\theta}} ) $ \qquad\qquad $^2 k\log(n) + \sum_j \log(c_j)$ \qquad\qquad $^3 - \frac{p}{2} \log(3) + \sum_i^p (\frac{1}{2}\log(I_{ii}) + \log(|\hat{\theta}_i|))$ \
		\end{tabular}
    \caption{Same as \cref{tab:cc_results}, but for the Pantheon+ supernovae data. 
    The Friedmann equation is again not the optimal fit to the data, although the difference in description length
    compared to the top ESR function is smaller
    than for cosmic chronometers,
     due to the greater
    increased constraining power of 
    the SNe.
    }
    \label{tab:panth_results}
    \end{center}
\end{table*}

\subsection{ESR on mock data}
\label{sec:mock}

We have observed that current cosmic chronometer and supernovae data are not sufficiently constraining to prefer $\Lambda$CDM over other expansion histories with smaller functional and parametric complexities. In this section we investigate what one would require in order to recover $\Lambda$CDM from a cosmic chronometer dataset by analysing mock catalogues.

We begin by drawing $N_{\rm cc}=32,000$ redshifts randomly from the range used in \cref{sec:cosmic_chronometers} (corresponding to a dataset which is 100 times larger than used here), drawing from a uniform distribution in $\log\left(1+z\right)$. For each redshift, we compute the ``true'' Hubble expansion rate using \cref{eq:Friedmann_LCDM} and assuming the Planck 2018 cosmology \cite{Planck18_Parameters}. To convert this to an ``observed'' Hubble parameter, we assume that all measurements have the same fractional uncertainty, $\sigma_H / H$, and draw these observations from a Gaussian with this width and a mean corresponding to the ``true'' value. Since the mean fractional uncertainty of the data used in \cref{sec:cosmic_chronometers} is approximately 20 per cent, we consider four different cases $\sigma_H / H \in \{ 0.01, 0.05, 0.10, 0.20 \}$ to investigate the effect of improved precision. We then run the ESR algorithm on these data and rank the functions according to the MDL principle.

Once we have run the analysis at fixed $N_{\rm cc}$, we wish to convert our results to arbitrary $N_{\rm cc}$. Assuming that one obtains the same parameters if one changed $N_{\rm cc}$, the only terms that dependent on $N_{\rm cc}$ in \cref{eq:length_final} are those containing $\log \mathcal{L}$ and $I_{ii}$. Both of these variables scale linearly with $N_{\rm cc}$, allowing for a trivial transformation. Note that this introduces a term $p/2 \log N_{\rm cc}$ into the description length, which is reminiscent of one of the terms in the BIC.

\begin{figure}
    \centering
    \includegraphics[width=\columnwidth]{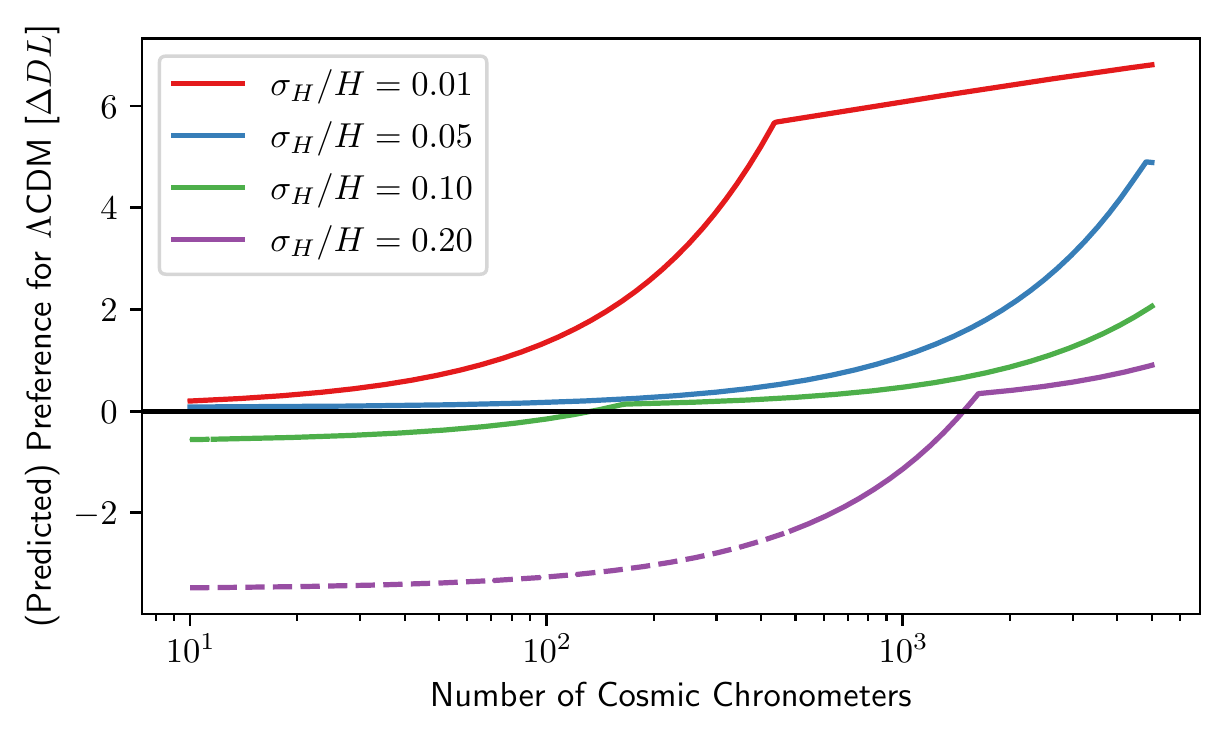}
    \caption{Change in description length between a $\Lambda$CDM expansion history and the highest-ranked non-$\Lambda$CDM function when applied to mock $\Lambda$CDM data as a function of number of cosmic chronometers. Any point greater than zero finds $\Lambda$CDM to be the best function and dashed lines indicate analyses where $\Lambda$CDM is not in the top two functions. We run the analysis assuming different fractional errors, $\sigma_H / H$, on measurements of the Hubble parameter, $H$, which for the observed data used in this work is approximately 0.2.
    }
    \label{fig:CC_mock}
\end{figure}

For cases where $\Lambda$CDM ranks as the highest function, we compute the difference in description length between \cref{eq:Friedmann_LCDM} and the second best function. Otherwise, we compute the difference between $\Lambda$CDM and the highest ranked function. We plot our findings as a function of $N_{\rm cc}$ for different $\sigma_H / H$ in \cref{fig:CC_mock}, where all points above zero are cases where we would find $\Lambda$CDM as the best fitting function. We indicate cases where $\Lambda$CDM does not lie in the top two functions by dashed lines.

We find that with current observational uncertainties, one would require $\sim 10^3$ measurements of $H$ in this redshift range to have sufficient constraining power for $\Lambda$CDM to be the best function. It is therefore no surprise that \cref{eq:Friedmann_LCDM} performed so poorly on the observed data, where we only have 32 data points. However, if one could reduce the observational uncertainties by a factor of two, then one only requires a few hundred cosmic chronometers for $\Lambda$CDM to be recovered. The forecast is even more promising if one could measure $H$ to within 1 or 5 per cent, since in both cases we see that $\Lambda$CDM has the MDL, even if one had only ten cosmic chronometers. These findings therefore suggest that reducing the observational uncertainties on cosmic chronometer measurements is more important than increasing the sample size if one wishes to discriminate between different expansion histories. 

\section{Discussion}
\label{sec:disc}

\subsection{Limitations of ESR}
\label{sec:limitations}
Before we proceed with a more detailed comparison with other, well-established methods of SR, it is useful to briefly recap the limitations of our method. Most importantly, it is limited to functions of lower complexity than other methods; due to its thoroughness it cannot reach complexities $\gtrsim$10 for reasonable operator basis sets.
Hence, we only consider univariate applications in this paper. The larger the basis set, the larger the number of possible combinations of functions at a fixed complexity and hence the greater the runtime and memory requirement of the algorithm. An important part of this is the procedure for finding duplicate functions, which is essential for making ESR efficient. The procedure is however imperfect: we cannot identify every duplicate and hence must optimise the parameters of more than just the unique equations.
Finally, the method is not optimised to find the best functions (the functions of lowest description length or those on or close to the Pareto front). While we are guaranteed to find them (provided that we find the global optima of the model parameters), it is only after an exhaustive scan is undertaken at each level of complexity. This contrasts (as described below) with other deterministic methods which use heuristics to increase the probability of finding good functions without a complete search.

\subsection{Comparison with other SR algorithms}
\label{sec:SR_comparison}

The ESR algorithm systematically searches through all possible equations which can be generated from a given set of basis functions and represented by trees with a fixed number of nodes. This is in contrast to traditional SR methods which do not consider all possible equations but rather attempt to search stochastically through the space of functions.
Since a key goal of SR is interpretability, one typically does not want to produce highly complex functions, making an exhaustive search of function space feasible.
We have found that the $L(D)$ minimum for both the cosmic chronometer and Pantheon+ datasets is at complexity 5, showing that even real datasets with up to 1590 points can have optimal functional representations well within the scope of ESR according to MDL.

In the event that one did desire a symbolic expression with high complexity, then one would have to utilise one of these probabilistic methods. The necessity (or not) of this does not need to be estimated \textit{a priori} but can be determined empirically from ESR: if a minimum of the description length is reached within the complexity range accessible to ESR then it is unlikely that further improvement is possible at higher complexity. (In principle there may be local minima in L(D), but we do not find this to be a common occurrence.) Other algorithms have a generally unknown probability of failure (not finding the true optimal function in a reasonable amount of time) that is an unknown function of their hyperparameters, and thus it is important to be able to benchmark these methods. By calculating the true Pareto front explicitly at low complexity, ESR allows the user to evaluate different stochastic methods on their dataset; if the SR algorithm cannot find the true MDL function or Pareto front at low complexity, results at higher complexity are unlikely to be trustworthy. Similarly, once one has chosen a SR algorithm, the results of applying ESR to the data could be useful in hyperparameter optimisation or in genetic programming by initialising with a starting population of well-fitting functions.

Usually one only discusses the optimal solution or set of Pareto-optimal solutions when considering SR, as this is what
traditional SR algorithms are designed to find.
Although, in principle, genetic programming can provide the full population which it has found and its evolutionary history, common implementations (e.g. \textsc{PySR}) discard functions which do not lie on the Pareto front. A clear advantage of ESR is the ability to find all functions and their corresponding likelihoods and description lengths. Not only does this allow us to quantify the uncertainty on our prediction in a rigorous and complete way---particularly useful when extrapolating beyond the range of the training set---but we can now propagate uncertainties on the functional form of equations alongside the uncertainties on their parameters when making new predictions.
This can be achieved by performing functional integrals, weighting each function by the exponential of the negative description length. Such an analysis could be performed using equations generated in stochastic searches, although in that case one would not have an exhaustive set and hence may be missing low-description-length solutions.

We reiterate that, for our case studies, the equations given in \cref{tab:cc_results,tab:panth_results} do not necessarily give the ``true'' cosmological expansion rate, but are merely the most economical descriptions of the data. Having an exponent of $1+z$ may be theoretically unfavourable, however the ``function prior'' term of ESR ($k\log(n)$) depends only on the number of nodes and basis operators and not on how ``physically reasonable'' they are. One may therefore want to investigate the impact on different priors on function space, e.g. by applying
theoretical priors in post-processing. An advantage of ESR over traditional algorithms is that, given the ESR algorithm generates and fits all possible equations, one only needs to run the optimisation once; one can easily reweight or remove equations according to different criteria without having to rerun the equation search or parameter optimisation. Of course, if one decided to use a different loss then one would have to rerun the parameter optimisation.

In \cref{sec:MDL} we derived an expression (\cref{eq:length_final}) from the description length (\cref{eq:length}) of a model which naturally combines the accuracy with the parametric and structural complexity of a function into a single summary, which one should optimise to determine which equation is preferred by the data. This statistic incorporates the features one would expect should be in such a criterion: models with lower accuracy, more parameters and more nodes are all punished. Other SR codes tend to consider only a subset of these features in their selection procedure. For example, \textsc{PySR} considers the derivative of the logarithm of the loss with respect to the complexity; however \cref{eq:length_final} suggests differentiating with respect to the logarithm of complexity is more appropriate.
In \textsc{aifeynman} the description lengths of real numbers are calculated with a free parameter precision, 1/$\Delta_i$, which scales the parameter part of $L\left(D\right)$ and hence leads to essentially arbitrary values. Their choice for 1/$\Delta_i$, $2^{-30}$, greatly penalises real numbers relative to integers or rationals. It is better motivated to select the precisions per parameter according to the Fisher information. The MDL principle stands alone from the function generation part of ESR as a model selection algorithm, and we recommend that an implementation along the lines of \cref{eq:length_final} be incorporated into all SR algorithms.

This is not however to say that the quantification of model complexity has not been extensively studied in the literature (e.g. \cite{Smits,Kommenda,Chen,Bomarito}). In particular, a potential limitation of MDL is that it uses purely structural (genotypic) as opposed to behavioural (phenotypic) complexity and therefore allows for highly nonlinear behaviour such as discontinuous derivatives \cite{ESR_on_RAR}. Incorporating a penalisation term for the order of nonlinearity \cite{Order_nonlinearity} could rectify this in cases where smoothness is important. We note that MDL was first applied to genetic programming by \cite{Iba}, where it was used to quantify the goodness-of-fit of decision trees in a classification context.

Deterministic methods have also been previously designed for symbolic regression (e.g. \cite{Worm, Kammerer_2021, Dome}). A common strategy is to start from a very simple set of basis functions and then consider all possible ways of adding to the functions' trees to increase the complexity. To expedite the process, functions to expand may be prioritised according to their quality (accuracy and/or simplicity). This affords a deterministic search through function space towards higher complexity along pathways of well-fitting functions, controlled by hyperparameters such as the relative weighting of accuracy and simplicity \cite{Kammerer_2021}. If a given route comes to appear unpromising, expanding it further may acquire lower priority than trying expansions in a different region of function space instead.
Thus, given enough time these algorithms would explore all possible functions, analogously to ESR. Normally the goal for these deterministic methods---and their advantage---is to reach good functions quickly and then terminate the search, rather than evaluate the space systematically as in the case of ESR. We note, however, that there will always exist cases where the heuristics used to expedite the search are counterproductive. Exhaustive scanning obviously does not suffer from this ``no free lunch'' \cite{Wolpert} problem. Further advantages of ESR are that it is not iterative so is much more efficiently parallelisable, and that it does not have reduced probability of evaluating functions qualitatively dissimilar to high-quality functions found previously. This helps to discover substantially novel solutions and prevents the algorithm becoming stuck trying to improve sub-optimal solutions with noisy data (a problem reported by \cite{Kammerer_2021}). We also consider different functions to \cite{Kammerer_2021}: they limit the amount of operator nesting and increase complexity through the addition of new terms, whereas we can reach higher complexity through either addition of new terms or nesting.

\subsection{Benchmarking ESR against other algorithms}

For a more quantitative comparison between SR algorithms, we use the univariate benchmark dataset \texttt{feynman\_I\_6\_2a} \cite{Olson_2017,Romano_2021} from the Penn Machine Learning Benchmarks\footnote{\url{https://github.com/EpistasisLab/pmlb}} dataset, as used in the SRBench\footnote{\url{https://cavalab.org/srbench/competition-2022/}} competition. This dataset consists of $10^5$ data points generated from an unknown function with zero observational error. We therefore compare the algorithms by the mean square error (MSE) of the functions they produce, rather than the MDL since without uncertainties there is no likelihood. We compare ESR against
\textsc{PySR} \cite{cranmer2020discovering,pysr},
\textsc{DataModeler} \cite{DM},
\textsc{ffx} \cite{FFX}
and \textsc{QLattice} \cite{QLattice}.
We run \textsc{PySR} and \textsc{DataModeler} with the same basis functions as in \cref{sec:case studies} for 10 hours on 24 cores (dual 12-core Xeon 2.2 GHz CPUs), although we find the same results if we run the algorithms for only 5 hours.
We use the default settings for all codes (including parameter optimisation strategies), but with batching turned on for \textsc{PySR} (as recommended for this number of points) and we combine results from using a maximum edge number of three, four or ten for \textsc{QLattice}.
Since \textsc{DataModeler} uses a different definition of complexity to us, we output the Pareto layers containing at least 20\% of best functions it finds and recompute complexity for these to find the best Pareto front according to our definition. We also apply this procedure to all models produced by \textsc{QLattice}.

\begin{figure}
    \centering
    \includegraphics[width=\columnwidth]{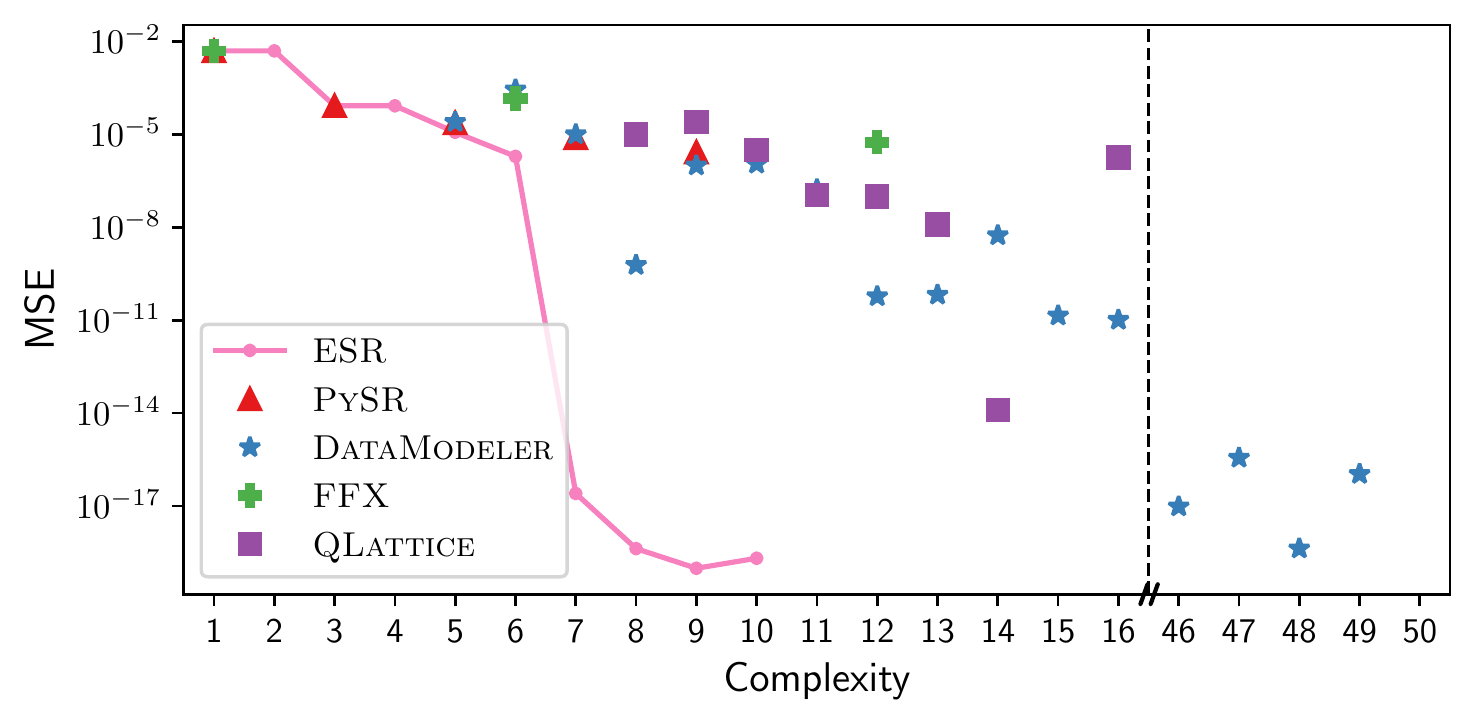}
    \caption{Pareto front of solutions found by different SR algorithms for the \texttt{feynman\_I\_6\_2a} benchmark dataset. By construction, ESR finds the optimum solution at a given complexity, whereas other methods have an unknown (high) probability of failing to find this. The vertical dashed line indicates a break in the x-axis.}
    \label{fig:benchmark_pareto}
\end{figure}

The Pareto front found by the different methods is plotted in \cref{fig:benchmark_pareto}. By construction, ESR has found the best solution at each complexity, whereas the other algorithms are able to match this at low complexity, but typically converge to subsidiary optima at higher complexity, with MSEs which can be many orders of magnitude worse than found with ESR. 
The best-fit ESR function at complexity 7 is
\begin{equation}
    y = \theta_1 \theta_0^{x^2},
\end{equation}
with $\bm{\theta}=(0.60653063, 0.39894231)$, and all functions on the Pareto front at higher complexity also take this form.
Upon closer inspection, one finds that $\theta_0 \approx 1 / \sqrt{e}$ and $\theta_1 \approx 1 / \sqrt{2 \pi}$ (with differences of $\mathcal{O}(10^{-8})$), providing the insight to try a standard Gaussian. This yields MSE = $3\times10^{-33}$ and hence we conclude that we have discovered the generating function. We note that the only reason we do not match this MSE is due to the finite precision of the numerical optimisers, but this is a customisable property of the optimisers and not a limitation of ESR itself.

\textsc{DataModeler} was able to produce a comparable MSE to the best found by ESR, but only at a much higher complexity. Examination of this function shows that it is not a Gaussian and hence provides no insight into the true generating function: it is merely a complicated fitting function. We verified that no other algorithm returned a function with the correct form and that re-optimising parameters using the method described in \cref{sec:param_opt} produces insignificant changes to the MSE.

For ESR, we find that we can use slightly relaxed settings compared to the defaults of \cref{sec:param_opt}, $N_{\rm iter}=5$, $N_{\rm conv}=2$ and a reduced function walltime limit of $1{\rm \, s}$, and achieve practically identical results. Including function generation, this had a run time of $\sim$150 CPU hours, which is significantly less than we allocated for \textsc{DataModeler} and \textsc{PySR} (240 CPU hours). If instead one used the precomputed functions, then one would only require 33 CPU hours with these settings. As mentioned above, halving the run time for \textsc{DataModeler} and \textsc{PySR} does not change the results. This indicates that either these algorithms have converged (which is problematic as they have converged to the wrong answer) or that more time is required to converge, which is already longer than ESR.

This benchmark test highlights the advantages of an exhaustive search; if one only ran the non-exhaustive algorithms, one would have no way of knowing that there exists such a simple function which can fit the data very accurately. Although one could run a hyperparameter search to improve the results of the other algorithms (adding significantly to the runtime), it would remain unclear whether the true optimum had been found.

We reach similar conclusions using noisy data. We ran \textsc{PySR} with the same configuration as above to determine the Pareto front for the cosmic chronometer dataset (\cref{sec:cosmic_chronometers}), attempting to find the same function as ESR and using the same likelihood (\cref{eq:cc_likelihood}). \textsc{PySR} is able to obtain the true Pareto front up to and including complexity 6, but achieves little improvement beyond that. ESR is able to find solutions with a $\log\mathcal{L}$ of up to 2.0 smaller than the \textsc{PySR} function at complexity 7-10. Again, we find practically identical results if we run \textsc{PySR} for 5 or 10 hours, suggesting it has practically converged.

\subsection{Future directions}

We now briefly discuss possible future developments of the algorithm we propose here. To begin with, for simplicity, we focused on univariate functions but, inevitably, we will want to consider multivariate functions. Such an extension is conceptually straightforward although will face the problem of increased number of functions at a given complexity. 

In the future one may want to consider higher complexity functions. The main bottleneck is the task of finding duplicates; we have used the \textsc{sympy} package which requires a large amount of memory once one exceeds $\mathcal{O}\left(10^6\right)$ functions. Furthermore, \textsc{sympy} fails to find duplicates in many situations. It makes sense, then to develop a simpler, custom symbolic manipulation package with a particular focus on improving the duplicate finding procedure along the lines of \cite{Kammerer_2021,Burlacu_2021}.

Previous deterministic SR methods reach higher complexity without a complete scan of function space. ESR could provide a way to increase the efficiency of these searches by initialising them at much higher complexity where a greater range of model behaviour is possible. This would reduce the risk of failing to find high-quality functions that are qualitatively different from the best-fitting very simple ones. Another possibility is to use the complete knowledge of functional parameter space achieved by ESR to assign probabilities to the possible transformations in the priority queue for expanding them towards higher complexity. This would allow both the functions to expand and the transformations to apply to be learnt from the data, which should expedite the discovery of good new functions.

Another potential improvement is in the method for removing duplicate equations. While were able to cull 97 per cent of possible functions by considering the topology of tree representations,  we find that only up to 26 per cent of the remaining functions are unique. One may therefore wonder whether there is a more economic notation one could exploit with fewer redundancies.

Furthermore, when we optimise parameters we do not attempt preferentially to find integers or rational numbers over floats. Since such parameters could be encoded with fewer nats than floats, there could exist functions with shorter description lengths than the ones we consider. One way of doing this is if one could ``snap'' floating-point parameters to integers. The Fisher matrix calculated as part of the parameter codelength provides a natural way to do this by determining which integers or rationals are close enough to the best-fit float not to degrade the likelihood significantly. One may also wish to prioritise certain irrational numbers which commonly appear in physics (e.g. $\pi$, $e$ or $\sqrt{2}$), e.g. by defining corresponding custom nullary operators.

Our approach has privileged no class of function, which may be thought of as a uniform prior over functional parameter space (given the predefined operator set and the $k\log(n)$ penalisation term). However, it may be possible to exploit symmetries in functions to reduce the description length, or to prefer Hamiltonian or Lagrangian systems. This would favour equations more likely to be found in physics and improve convergence on datasets possessing such features.

Finally, in this work we have only considered the maximum likelihood values of parameters. Although the Fisher matrix term of \cref{eq:length_final} prefers parameters which achieve a good fit to data across a wider range of parameter space, it may be preferable to consider a more Bayesian approach. This would involve using the Bayesian evidence instead of the maximum likelihood in \cref{eq:length_final}, and thus would marginalise over the uncertainties in the parameters; we leave a complete analysis for future work. We note, however, that in \cite{ESR_on_RAR} we show how explicit Bayesian inference using the lowest description length functions may be used to determine the posterior predictive distributions of physically interesting quantities and hence evaluate the probabilities of competing classes of models.

\section{Conclusion}
\label{sec:conc}

Symbolic Regression (SR) is a class of inherently interpretable machine-learning methods which learn the analytic expressions that accurately describe data. The conventional approach to this problem involves a stochastic search through parameter space, through e.g. genetic programming, reinforcement learning or by searching for symmetries in the data, to produce a Pareto front of viable solutions which balance simplicity with accuracy. Such techniques have a generally unknown probability of failing to obtain the best solution, and little effort has gone into comparing the objectives of simplicity and accuracy to obtain the optimal equation.

In this work we have introduced a novel approach to SR---\emph{Exhaustive Symbolic Regression} (ESR)---which aims to address both of these issues. First, motivated by the relative simplicity of the functions which tend to be selected by SR algorithms, we eliminate the risk of failing to obtain the optimal solution by explicitly considering all possible functions at a given complexity (defined as the number of nodes in the tree representation of a function) made from a given operator set. Although this problem scales exponentially with complexity, by considering the topology of valid trees and applying simplification procedures to identify duplicate functions, we find that we can reduce the number of functions one needs to fit to the data to a feasible number, which is a factor of $\sim1400$ smaller than the na\"{i}ve expectation. Second, we derive a statistic (\cref{eq:length_final}) which combines the likelihood and Fisher matrix with the parametric and structural complexities of a function into a single number describing the amount of information that must be communicated to specify the data with the help of the function. The minimum description length principle asserts that the function minimising this number provides the optimal description of the data. In this way, we collapse the Pareto front into a one-dimensional ranking and thus remove the ambiguity associated with model selection in dual-objective SR algorithms (although of course the full Pareto front may be considered in ESR if desired). As a demonstration of our method, we reconstruct the (square of the) Hubble parameter as a function of redshift from both cosmic chronometers and the Pantheon+ sample of Type Ia supernovae.

 We publicly release our code and the equations generated in this work \cite{ESR_data}. Although we considered one set of basis functions in our analysis, this can be trivially modified to generate the appropriate function set for a given analysis. For example, in our companion paper \cite{ESR_on_RAR} we apply our method to learn the functional form of the Radial Acceleration Relation of galaxy dynamics, and, motivated by previous fitting formulae, add $\powsqr$, $\exp$ and $\sqrt{\cdot}$ to our basis set.
The equations generated in that paper are also made publicly available \cite{ESR_data}.

Automated discovery of symbolic expressions from data provides easily understood and readily usable descriptions of complex physical processes, and can even discover fundamental physical laws. It is therefore imperative that the model generation and selection algorithms of SR score equations in a rigorous and well-motivated way.
By meeting this challenge, the ESR framework improves the prospects of SR to
become a key method for data analysis.

\section*{Data availability}

We make available the ESR code (\url{https://github.com/DeaglanBartlett/ESR}) \,and function sets generated in this work \cite{ESR_data}.
The cosmic chronometer data used in \cref{sec:cosmic_chronometers} can be found in Table 1 of Moresco et al. \cite{Moresco_2022}. The supernova data described in \cref{sec:supernovae} was taken from the publicly released Pantheon+ catalogue (\url{https://github.com/PantheonPlusSH0ES/DataRelease}). Other data may be shared on request to the corresponding authors.

\section*{Acknowledgements}

We thank D. Bacon, A. Constantin, M. Cranmer, M. Figueiredo, G. Gregori, T. Harvey, J. Jasche, L. Kammerer, M. Kotanchek, G. Kronberger, A. Lukas, B. Wandelt and T. Yasin for useful inputs and discussion. We thank J. Patterson for smoothly running the Glamdring Cluster hosted by the University of Oxford.
DJB is supported by the Simons Collaboration on ``Learning the Universe'' and was supported by STFC and Oriel College, Oxford.
HD is supported by a Royal Society University Research Fellowship (grant no. 211046).
This project has received funding from the European Research Council (ERC) under the European Union's Horizon 2020 research and innovation programme (grant agreement No 693024), STFC and the Beecroft Trust.
This work used the DiRAC Complexity and DiRAC@Durham facilities, operated by the University of Leicester IT Services and Institute for Computational Cosmology, which form part of the STFC DiRAC HPC Facility (www.dirac.ac.uk). This equipment is funded by BIS National E-Infrastructure capital grants ST/K000373/1, ST/P002293/1, ST/R002371/1 and ST/S002502/1, STFC DiRAC Operations grant ST/K0003259/1, and Durham University and STFC operations grant ST/R000832/1. DiRAC is part of the National E-Infrastructure.
This work used the
\textsc{astropy} \cite{Astropy},
\textsc{matplotlib} \cite{Matplotlib},
\textsc{mpi4py} \cite{mpi4py}, 
\textsc{mpmath} \cite{mpmath}, 
\textsc{networkx} \cite{networkx},
\textsc{numdifftools} \cite{numdifftools},
\textsc{numpy} \cite{Numpy},
\textsc{pandas} \cite{reback2020pandas,mckinney-proc-scipy-2010},
\textsc{scipy} \cite{Scipy} and
\textsc{sympy} \cite{Sympy} packages.
For the purpose of open access, the authors have applied a Creative Commons Attribution (CC BY) licence to any Author Accepted Manuscript version arising.

\bibliographystyle{IEEEtran}
\bibliography{references_IEEE}

\label{lastpage}
\end{document}